\documentclass{egpubl}

\STAR                   
 \electronicVersion 

\ifpdf \usepackage[pdftex]{graphicx} \pdfcompresslevel=9
\else \usepackage[dvips]{graphicx} \fi

\PrintedOrElectronic

\usepackage{t1enc,dfadobe}

\usepackage{egweblnk}
\usepackage{cite}

\usepackage{amssymb}	
\usepackage{dsfont}

\title[Curve Reconstruction Benchmark]%
      {2D Points Curve Reconstruction Survey and Benchmark}

\author[S. Ohrhallinger \& J. Peethambaran \& A. Parakkat \& T. Dey \& R. Muthuganapathy]
	{S. Ohrhallinger$^1$
	and J. Peethambaran$^2$
	and A.\,D. Parakkat$^3$
	and T.\,K. Dey$^4$
	and R. Muthuganapathy$^5$
        \\
         $^1$Institute of Visual Computing and Human-Centered Technology, TU Wien, Austria\\
	$^2$Department of Math and Computing Science, Saint Mary’s University, Halifax, Canada\\
	$^3$Department of Computer Science and Engineering, Indian Institute of Technology Guwahati, India\\
	$^4$Department of Computer Science, Purdue University, Indiana, USA\\
	$^5$Department of Engineering Design, Indian Institute of Technology Madras, India
       }

\usepackage{overpic}
\usepackage{amssymb}
\usepackage{amsmath}
\usepackage{graphicx}
\usepackage{wrapfig}
\usepackage{float} 
\usepackage[format=plain,
            labelfont=it,
            textfont=it]{caption} 
\usepackage{enumitem} 
\usepackage{microtype}
\usepackage{algorithm}
\usepackage[noend]{algpseudocode}
\usepackage{subcaption}

\newcommand{\hidden}[1]{{}} 

\newcommand{\ADDRESSED}[1]{{}}

\let\vec=\mathbf

\newcommand{\lfs}{\mbox{lfs}}
\newtheorem{definition}{Definition}

\begin{document}

 \teaser{
  \includegraphics[width=\linewidth]{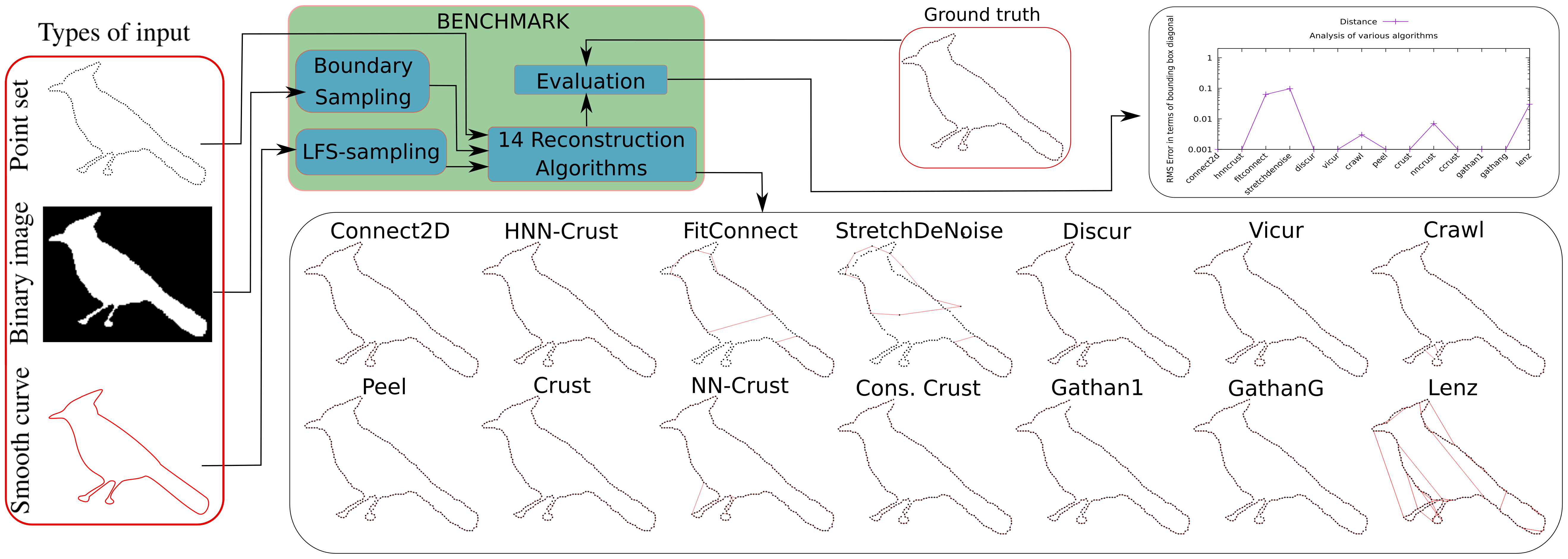}
  \centering
   \caption{We survey 36 curve reconstruction algorithms and compare 14 of these, with quantitative and qualitative analysis. As inputs we take unorganized points, samples on the boundary of binary images or smooth curves, and evaluate with ground truth.}
 \label{fig:teaser}
 }
\maketitle
\begin{abstract}
Curve reconstruction from unstructured points in a plane is a fundamental problem with many applications that has generated research interest for decades.
Involved aspects like handling open, sharp, multiple and non-manifold outlines, run-time and provability as well as potential extension to 3D for surface reconstruction have led to many different algorithms.
We survey the literature on 2D curve reconstruction and then present an open-sourced benchmark for the experimental study.
Our unprecedented evaluation on a selected set of planar curve reconstruction algorithms aims to give an overview of both quantitative analysis and qualitative aspects for helping users to select the right algorithm for specific problems in the field.
Our benchmark framework is available online to permit reproducing the results, and easy integration of new algorithms.

\end{abstract}

\section{Introduction}

Given a finite set of points $P$ sampled from a planar curve $\Sigma$, recovering a polygonal approximation to $\Sigma$ from $P$ is generally known as curve reconstruction. Reconstruction of curves is a fundamental task in many applications such as reverse engineering of geometric models, outline reconstruction from feature points in medical imaging systems, and facial feature detection in the face recognition algorithms ~\cite{ahlvers2005model}, among others, and an interesting problem by itself. Despite over three decades of tremendous research effort in the computational geometry, computer vision and graphics research communities, specific cases are still open in curve reconstruction, and there is no algorithm that would succeed on all types of problems. 
Recent research trends, however, address specific aspects of reconstruction such as improved sampling conditions ~\cite{ohrhallinger2016hnn}, reconstructing from fewer number of samples and curves with sharp corners ~\cite{url:connect2d}, reconstruction from unstructured and noisy point clouds~\cite{ohrhallinger2018fitconnect}, a unified framework for reconstruction ~\cite{METHIRUMANGALATH201590}, incremental labeling techniques for curve extraction \cite{doi:10.1111/cgf.13589}, and applications of curve reconstruction to hand drawn sketches ~\cite{parakkat2016crawl}.

A major hurdle to the ongoing efforts in designing new algorithms for curve reconstruction is the lack of a framework that provides a set of standard tools, algorithms and data for comparing and evaluating various reconstruction algorithms. Currently, algorithmic evaluations in this domain heavily rely on visual comparison. A meaningful practice in the empirical evaluation of reconstruction techniques is to compare the reconstructed results against the ground truth curves using error norms such as Hausdorff distance or $L2$-error norms. However, each research group has its own data set or generates the input data by sampling shapes from images. Such practices make it extremely difficult for the researchers and practitioners from other fields to assess the performance of different curve reconstruction techniques, and to conclusively determine a suitable algorithm for their scientific studies or applications. Furthermore, in most of the cases the algorithmic choices for the comparison are made based on the availability of implementations in the public domain.

To address these challenges, we have set up a benchmark framework for 2D curve reconstruction (https://gitlab.com/stefango74/curve-benchmark).
It consists of a set of \textbf{\emph{fifteen}} curve reconstruction techniques including the recent ones and a few support tools including a curve sampler that generates samples from smooth curves based on the $\epsilon$-sampling \cite{amenta98curve} criterion. Additionally, the benchmark provides a set of commonly used input data along with the ground truth curves for the experimental evaluation of algorithms in this domain. A set of newly generated input data that exhibits diverse features and is suitable for empirical studies is also included in the benchmark. Finally, we provide curve reconstruction evaluation criteria and features.

Besides presenting the curve reconstruction benchmark to the reader, this paper also covers principles and practices used in the curve reconstruction domain. We review the theoretical background, algorithms and their evolution, supporting tools, and evaluation criteria for curve reconstruction. Advantages and limitations of different methods are discussed. Apart from setting up the benchmark and reviewing various algorithms, the main contribution of this paper is an experimental assessment of the current state-of-the-art in the field with respect to standard error metrics such as Hausdorff distance, root mean square error (RMSE) and normal deviation. In the end, we delineate a few directions for future research.

\subsection{Contributions}
The main focus of this work is to review the available curve reconstruction literature and evaluate a competitive subset of curve reconstruction algorithms which take un-oriented and unorganized points as input and generate polygonal approximations to their underlying curves. We make the following key contributions.
\begin{itemize}
  \item \textbf{Algorithms Review} A comprehensive review of the curve reconstruction literature up to date.
  \item \textbf{Benchmark} A benchmark consisting of prominent curve reconstruction algorithms, supporting tools, existing data and new test data along with the ground truth.
  \item \textbf{Evaluation} A thorough performance evaluation study comprising the algorithms provided in the benchmark. The study helps in demonstrating how the benchmark can be utilized for selecting the right curve reconstruction algorithm for a specific problem.
\end{itemize}

\subsection{Related Work and Scope}

To our knowledge, two prior works exist in the literature and they both consider reconstruction of surfaces together with curves.
The first is a comprehensive book~\cite{dey2006curve} which describes the basic theory leading to the development of the $\epsilon$ sampling condition and relating it to algorithms from the {\scshape Crust} family with varying density before continuing to surface reconstruction algorithms, including noise and Morse theory there.
A later concise report~\cite{khanna2014survey} adds some faster local, visual- and optimization-based methods. An empirical evaluation of a few early curve reconstruction algorithms is presented in~\cite{althaus00curve}.

\begin{table*}
	\begin{center}
		\begin{tabular}{|l|r|r|r|r|r|r|r|r|r|r|}
			\hline
			Capabilities: & Param & \multicolumn{3}{c|}{Input} & \multicolumn{6}{c|}{Output}  \\
			\hline
			Algorithm & count & n.-u. & noise & outl. & manifold & open & mult. & sharp & guar. & time complexity \\
			\hline
			\textbf{Graph Based:} &  &  &  &  &  &  &  &  &  & \\
			$\alpha$-Shapes~\cite{edelsbrunner83alpha} & 1 & no & no & no & yes & no & yes & no & yes & O(n log n) \\ 
			$\beta$-skeleton~\cite{kirkpatrick85beta} & 1 & yes & no & no & no & no & yes & yes & no & O(n log n) \\
			$\gamma$-neighborhood~\cite{veltkamp19933d} & 2 & yes & no & no & no & no & yes & yes & no & O(n log n) \\
			EMST-based~\cite{figueiredo94curve} & 0 & no & no & no & yes & only & no & no & yes & O(n log n) \\ 
			Ball-pivoting~\cite{bernardini1997sampling} & 1 & no & no & no & yes & no & yes & no & yes & O(n log n) \\ 
			{\em r}-regular shapes~\cite{attali97regular} & 1 & no & no & no & no & no & yes & no & yes & O(n log n) \\ 
			Edge exchanging~\cite{ohrhallinger11operations} & 0 & yes & yes & no & yes & no & no & yes & no & NP \\
			{\scshape Connect2D}~\cite{ohrhallinger13connect2d} &0 & yes & yes & no & yes & no & no & yes & yes & O(n log n) \\ 
		         Shape-hull graph~\cite{PEETHAMBARAN201562} & 0 & yes & no & no & yes & no & no & yes & no & O(n log n) \\
			Voronoi Labeling~\cite{doi:10.1111/cgf.13589} & 0 & yes  & no  & yes & yes & no  & yes  & yes & yes & O(n log n)  \\ 
			{\scshape Crawl}~\cite{parakkat2016crawl} & 0 & yes & no & yes & no &  yes & yes & no & no & O(n log n) \\
			\hline
			\textbf{Feature Size Criteria:} & &  &  &  &  &  &  &  &  &  \\
			{\scshape Crust}~\cite{amenta98curve,gold99anticrust} & 0 & yes & no & no & yes & no & yes & no & yes & O(n log n) \\
			{\scshape NN-Crust}~\cite{dey99curve} & 0 & yes & no & no & yes & yes & yes & no & yes & O(n log n) \\
			{\scshape Cons. Crust}~\cite{dey99conservative} & 0 & yes & no & yes & no & yes & yes & no & yes & O(n log n) \\
			~\cite{lenz06curve} & 2 & yes & no & no & no & yes & no & yes & yes & O(n log n) \\
			~\cite{hiyoshi09optimize} & 0 & yes & no & no & yes & no & yes & no & yes & O($n^2$ log n) \\
			{\scshape HNN-Crust}~\cite{ohrhallinger2016hnn} & 0 & yes & no & no & yes & yes & yes & no & yes & O(n log n) \\
			\hline
			\textbf{Noisy Points Fitting:} & &  &  &  &  &  &  &  &  &  \\
			~\cite{lee2000curve} & 1 & yes & yes & yes & yes & yes & no & no & no & O($n^2$) \\
			~\cite{cheng2005curve} & 2 & yes & yes & no & yes & yes & yes & no & yes & O(n log n) \\ 
			{\scshape Robust HPR}~\cite{mehra2010visibility} & 5 & yes & yes & no & yes & yes & yes & yes & no & O(n log n) \\
			~\cite{rupniewski2014curve} & 1 & yes & yes & yes & yes & yes & yes & no & no & O(Mn) \\
			~\cite{wang2014robust} & 4 & yes & yes & yes & no & yes & yes & yes & no & O(d log d) \\
			{\scshape FitConnect}~\cite{ohrhallinger2018fitconnect} & 0 & yes & yes & yes & yes & yes & yes & yes & yes & O(n$k^2$) \\
			{\scshape StretchDenoise}~\cite{ohrhallinger2018stretchdenoise} & 0 & yes & yes & yes & yes & yes & yes & yes & yes & O(n$k^2$) \\
			\hline
			\textbf{Sharp Corners:} & &  &  &  &  &  &  &  &  &  \\
			~\cite{funke01curve} & 8 &  yes & no  & no  & no  & yes  & yes  & yes  & yes  & O(n log n) \\
			{\scshape Gathan}~\cite{dey01corners} & 1 & yes & no & no & no & yes & yes & yes & no & O(n log n) \\
			{\scshape GathanG}~\cite{dey02gathang} & 1 & yes & no & no & no & yes & yes & yes & yes & O(n log n) \\
			\hline
			\textbf{Traveling Salesman:} & &  &  &  &  &  &  &  &  &  \\
			~\cite{giesen99delaunay} & 0 & no & yes & no & yes & no & no & yes & yes & O(n log n) \\
			~\cite{althaus00polynomial} & 0 & yes & yes & no & yes & no & no & yes & yes & O(n log n) \\
			~\cite{arora1998polynomial} & 0 & yes & yes & no & yes & no & no & yes & yes & O(n (log n)$^{O(c)}$) \\ 
			Concorde solver~\cite{url:concorde} & 0 & yes & yes & no & yes & no & no & yes & TSP & NP \\
			\hline
			\textbf{Non-manifold:} & &  &  &  &  &  &  &  &  &  \\
			Opt. Transp.~\cite{degoes2011optimal} & 0 & yes & yes & yes & no & yes & yes & no & yes & O(n log n) \\
			{\scshape Peel}~\cite{parakkat2018peeling} & 2 & yes & yes & yes & no & yes & yes & no & yes & O(n$^{2}$) \\
			{\scshape ec-Shape}~\cite{METHIRUMANGALATH2017124} & 0 & yes & no & no & yes & no & no & no & yes & O(n log n) \\
			\hline
			\textbf{HVS Based:} & &  &  &  &  &  &  &  &  &  \\
			{\scshape DISCUR}~\cite{zeng08distance} & 0 & yes &no  &no  & yes & yes & yes & yes & yes & O(n log n)  \\
			{\scshape VICUR}~\cite{nguyen08vicur} & 4 & yes &no  &no  &yes  & yes & yes & yes & no & O(n log n)  \\
			\hline
		\end{tabular}
		\caption{Algorithms grouped by categories, with their input and output capabilities (n.-u. = nonuniform, guar. = sampling condition for manifold reconstruction)}
\label{table:algorithm-characteristics}
\end{center}
\end{table*}

In this survey, we look in detail at curve reconstruction and recent developments, e.g. in terms of theoretical guarantees.
In order to compare the algorithms and highlight their respective strengths, we have designed a benchmark for a comprehensive quantitative evaluation.
In Table~\ref{table:algorithm-characteristics} we compare the capabilities of 36 curve reconstruction algorithms (categorized according to type) w.r.t. to input point sets requirements and output piece-wise curves.

\subsection{Reconstruction Taxonomy}

\begin{figure}
	\centering
	\includegraphics[width=3in]{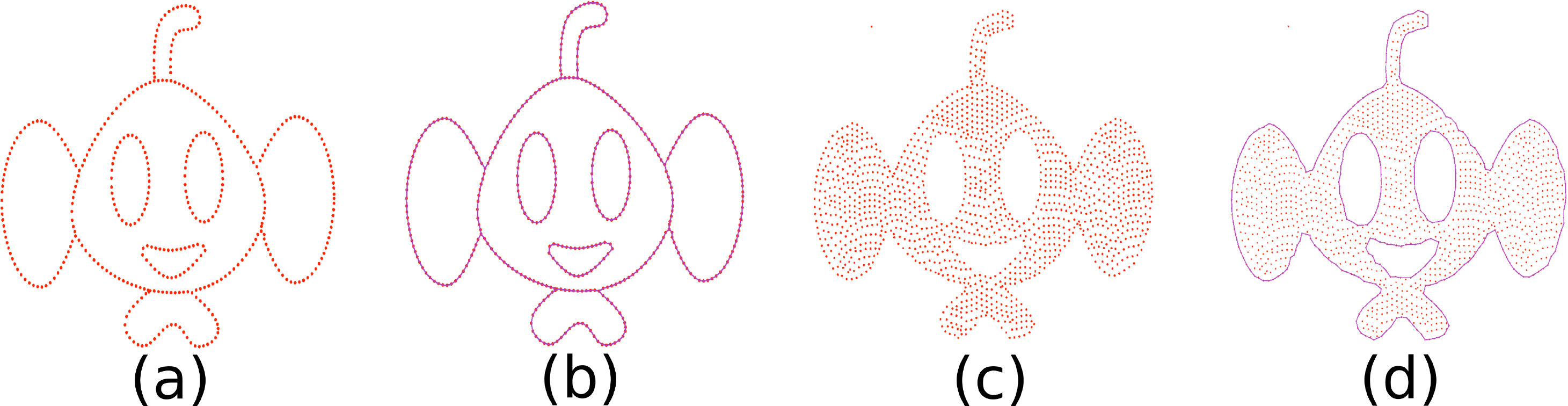}
	\caption{Two common types of inputs to the 2D reconstruction algorithms. (a) Boundary sample (b) Reconstructed curves (c) Area sample (d) Reconstruction from area samples.}
	\label{fig:input_types}
\end{figure}

\textbf{Boundary vs. Area Samples:} In general, there are two types of inputs to the polygonal reconstruction problems: \emph{boundary samples} and \emph{area samples}. Boundary samples consist of points sampled along a curve while area samples are sampled across an entire region including its boundaries as shown in Figure \ref{fig:input_types}. While curve reconstruction is a well defined problem, the reconstruction from area samples is ill-posed in nature~\cite{10.1007/BFb0054315, PEETHAMBARAN2015164}. The primary reason is a lack of precise mathematical definition for what constitutes the optimal approximation for the geometric shape of a point set with the points sampled from its interior. Furthermore, the shape perception from area samples is highly subjective since it often depends on a specific application context or human cognitive factors. A few unified algorithms \cite{METHIRUMANGALATH201590, DUCKHAM20083224, 5732742, THAYYIL2020101879, THAYYIL2021101953} that handle sampled boundaries as well as areas have also been proposed. Since there are numerous algorithms such as $\alpha$-shapes~\cite{edelsbrunner83alpha} that handle both the input types with reasonable accuracy, we have not included unified algorithms in our experimental study. In this work, we focus only on reconstruction from boundary samples.

\begin{figure}[ht]
	\centering
	\includegraphics[width=6cm]{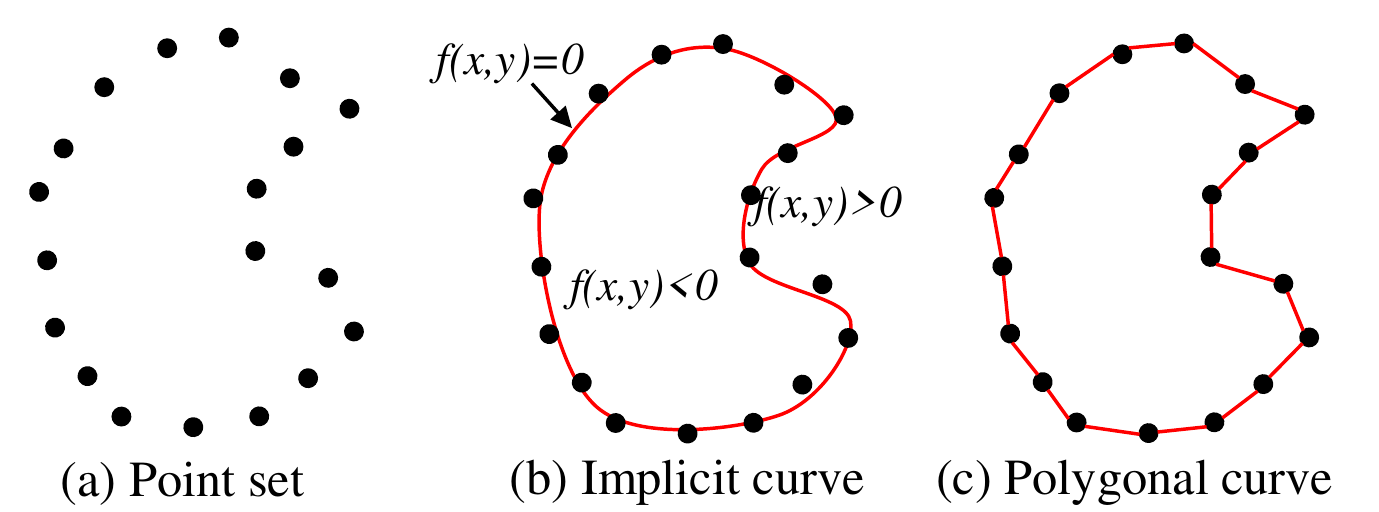}
	\caption{Illustration of implicit function in 2D and an example of explicitly reconstructed polygonal curve. }
	\label{fig:implicit}
\end{figure}

\textbf{Implicit Vs. Explicit:} Broadly, the reconstruction techniques can be grouped into implicit fitting and explicit reconstruction. Implicit methods attempt to define a smooth function $f \colon R^{2}\rightarrow R$ such that the zero level set of $f$ approximates the underlying curve in the input points as illustrated in Figure \ref{fig:implicit}. Explicit reconstruction deals with connecting the input points using edges or triangular faces subjected to certain geometric criteria, which results in a piece-wise linear approximation to the underlying curve. Implicit reconstruction algorithms utilize the orientation of the point sets to define the curve function. The orientation of the points are obtained through point normals or partially inferred through the segmentation of binary images (see Figure \ref{fig:teaser}). On the contrary, explicit reconstruction algorithms take un-oriented point sets. A vast majority of explicit methods interpolate all the input points including noisy data or outliers. As a result, explicit curve reconstruction algorithms are not robust to noise unless additional algorithmic criteria such as the ones in ~\cite{wang2014robust, ohrhallinger2018fitconnect, ohrhallinger2018stretchdenoise} to deal with the noise/outliers are included. Implicit methods work on noisy data. However, since the iso-curves are extracted using marching squares on quadtrees, an appropriate quadtree depth has to be determined and preset in the case of implicit methods. Highly detailed curves can be generated for larger depth values, however at the expense of increased computational time. While popular in 3D surface reconstruction, we have not found implementations or results for curve reconstruction, and so this category is not included in our comparison or evaluation. However, we have dedicated a section (Section \ref{sec:implicit-fitting}) to review the popular implicit fitting algorithms.

Categorizing the algorithms is a difficult task since some of them could fit with several aspects.
We propose the following taxonomy to subdivide them along their characteristics we deemed most important:

\begin{itemize}
	\item {\em Graph Based}: constructs a graph from the points and then filters the outline by some criterion
	\item {\em Feature Size Criteria}: differing approaches, but the required sampling density is proven in relation to local feature size
	\item {\em Noisy Points Fitting}: able to recover the original smooth curve from noisy samples
	\item {\em Sharp Corners}: can reconstruct angles $<90^\circ$ degrees instead of smoothing them over
	\item {\em Traveling Salesman}: minimizes total curve length
	\item {\em Non-manifold}: also handles (self-)intersections in curves
	\item {\em HVS Based}: inspired by the Gestalt laws on how Human eyes perceive visual elements
\end{itemize}

This taxonomy is also used to structure the descriptions of algorithms in Section~\ref{sec:algorithms}.

We examine the algorithms for the following properties:

\textbf{Input point set:}

\begin{itemize}
	\item {\em Non-uniformity}: no uniform point spacing required
	\item {\em Noise}: samples can be displaced from the original curve
	\item {\em Outliers}: additional points far from the curve are ignored
\end{itemize}

\textbf{Output piece-wise linear curve:}

\begin{itemize}
	\item {\em Manifoldness}: each vertex has $\le 2$ incident edges
	\item {\em Open curves}: end vertices (=holes in boundary) can exist
	\item {\em Multiply Connected}: curve has 2 or more components
	\item {\em Sharp corners}: angles $<90^\circ$ can be reconstructed
	\item {\em Guarantees/Conditions}: for successful reconstruction
	\item {\em Time complexity}: worst-case behavior of the algorithm.
\end{itemize}

In the following sub-section we detail challenges of these properties:


\subsection{Challenges for Reconstruction}


\begin{figure}[h]
\centering
\begin{subfigure}{.2\textwidth}
	\centering
	\includegraphics[width=3cm]{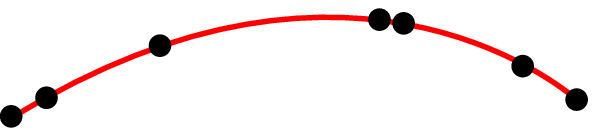}
	\caption{Non-uniform sampling}
	\label{subfig:nonuniform}
\end{subfigure}
\hspace{0.5cm}
\begin{subfigure}{.2\textwidth}
	\centering
	\includegraphics[width=3cm]{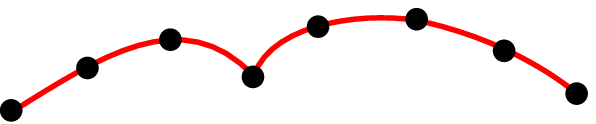}
	\caption{Sharp corner}
	\label{subfig:sharp}
\end{subfigure}
\begin{subfigure}{.2\textwidth}
	\centering
	\includegraphics[width=3cm]{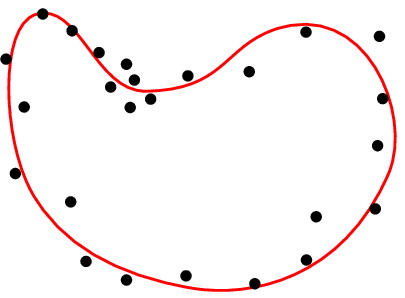}
	\caption{Noisy manifold}
	\label{subfig:manifold-noisy}
\end{subfigure}
\hspace{0.5cm}
\begin{subfigure}{.2\textwidth}
	\centering
	\includegraphics[width=3cm]{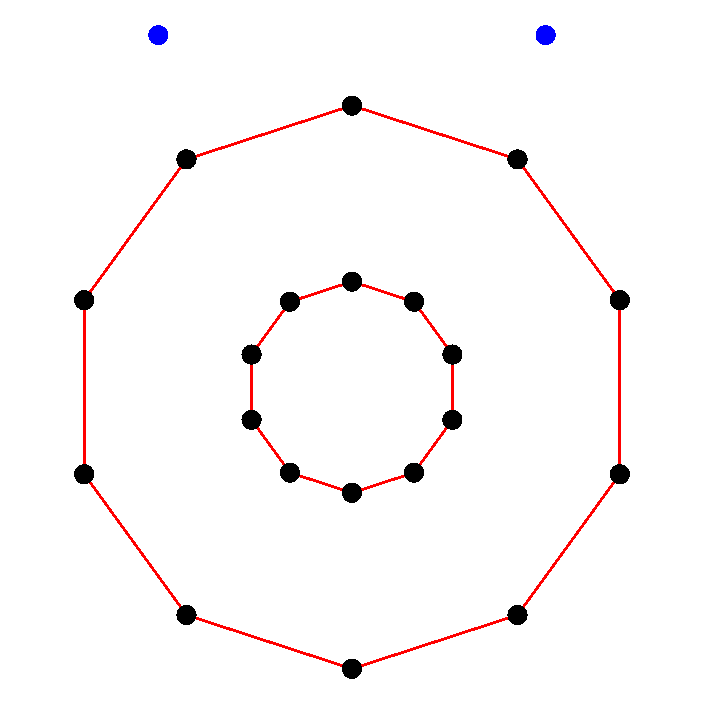}
	\caption{Multiply connected curve, outliers (blue)~\cite{ohrhallinger2018fitconnect}}
	\label{subfig:multiple-outlier}
\end{subfigure}
\begin{subfigure}{.2\textwidth}
	\centering
	\includegraphics[width=3cm]{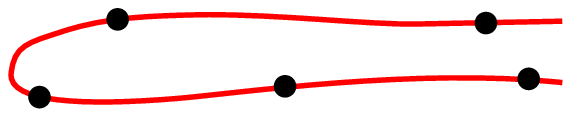}
	\caption{Sparse sampling}
	\label{subfig:sparse}
\end{subfigure}
\hspace{0.5cm}
\begin{subfigure}{.2\textwidth}
	\centering
	\includegraphics[width=3cm]{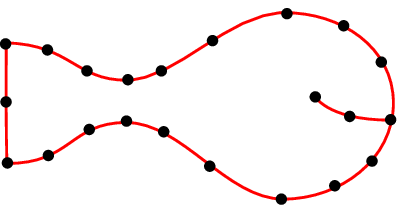}
	\caption{Non-manifold curve}
	\label{subfig:nonmanifold}
\end{subfigure}
	\caption{Various challenging input point configurations for curve reconstruction.}
	\label{fig:input-challenges}
\end{figure}

Here we give a list of challenging aspects in curve reconstruction (see Figure~\ref{fig:input-challenges}) - all of the input configurations can also be found combined:

\textbf{Non-uniformity}: Some algorithms require a globally uniform maximum spacing between samples. The disadvantage is that only features larger than that distance can be reconstructed. Algorithms which can reconstruct from non-uniform sampling (see Figure~\ref{subfig:nonuniform}) are then not limited to a specific absolute size for reconstructing features, but only restricted by too sparse sampling of features (see Figure~\ref{subfig:sparse}). A limitation can also be given for relative uniformity as a factor between spacings of adjacent samples.

\textbf{Noise}: When sensing data, such as silhouettes of objects, samples are usually perturbed by noise from measurement errors (see Figure~\ref{subfig:manifold-noisy}). Algorithms which strictly interpolate the input points will in the best case reconstruct a locally perturbed curve. Instead, points can be fitted in order to recover the original curve either by smoothing over the noise, or by denoising with knowledge/estimate of the local noise extent.

\textbf{Outliers}: Sensing data can also introduce erronous points far from the original curve. These are labeled outliers (see Figure~\ref{subfig:multiple-outlier}), should not be considered in the reconstruction, and must thus be classified and excluded beforehand.

\textbf{Manifoldness}: A boundary of an object is always a manifold curve, i.e. points are assigned at most two neighbors (see Figure~\ref{subfig:manifold-noisy}). Otherwise, the curve can become self-intersecting (see Figure~\ref{fig:input_types}a), which is useful for reconstructing drawings.

\textbf{Open curves}: Curves can be open if samples are missing from an object boundary, and still be manifold, also a collection of open curves, e.g. drawings (see Figure~\ref{subfig:nonuniform}) .

\textbf{Multiply Connected}: If boundaries of more than one object are to be reconstructed, these must not be interconnected in order to remain manifold. Holes in shapes are homotopy equivalent to that (see Figure~\ref{subfig:multiple-outlier}).

\textbf{Sharp corners}: Angles which are below 90 degrees are more difficult to reconstruct as it is more ambiguous which neighbors to connect if they are not in the opposite half-space of the other neighbor (see Figure~\ref{subfig:sharp}). Also, fitting algorithms tend to round off such sharp corners.

\textbf{Guarantees/Conditions}: It is very useful to know to which extent a curve can be reconstructed from a sampling. Guarantees can be given in terms of uniformly spaced sampling as a maximum global distance value or relative factor between neighbor points, a sampling condition in terms of the feature size (e.g. $\epsilon$-sampling), percentage of outliers, extent of noise range and statistical distribution, and as a distance of the reconstruction from the original curve.

\textbf{Time complexity}: Since points in a plane are mostly a small number such as a few thousands, optimization is not critical, but worst time complexity matters, as some algorithms have O($n^2$) or are not solvable in polynomial time.

\section{Preliminaries}
Many of the following definitions were introduced in this seminal paper~\cite{amenta98curve}.
We follow that up with an overview of proximity graphs which are used in many algorithms.


\subsection{Definitions and Notations}
Let $P$ be a set of $n$ points sampled from a simple closed planar curve $\Sigma$. The curve $\Sigma$ (if closed) is said to be \emph{convex} if the line segment between any two points on the curve falls in the interior, $I(\Sigma)$. Otherwise it is \emph{concave}. The curvature $\kappa$ at a point $p$ of $\Sigma$ is the rate of change of direction of the tangent line at $p$ with respect to arc length $s$. An inflection point ($IP$) on the curve is a point where $\kappa=0$ but $\kappa^{'}\neq0$. Concave subsets of a curve are characterized by the sign of the local curvature $\kappa$. Concave subsets exists between two inflection points and have a negative local curvature sign ($\kappa<0$) \cite{PEETHAMBARAN201562}.

\begin{figure}[h]
	\centering
	\includegraphics[width=4cm]{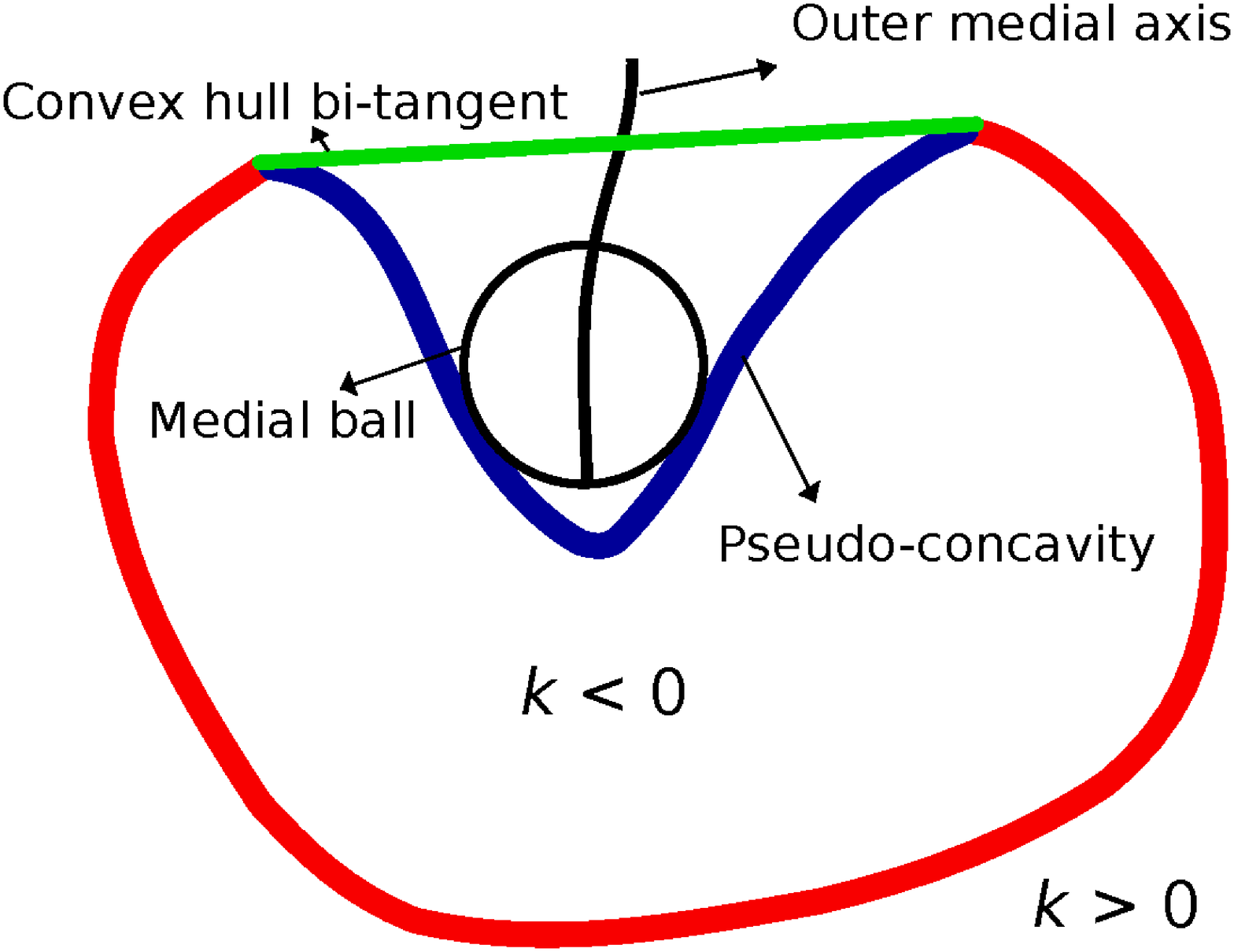}
	\caption{Illustration of pseudo-concave subsets of a simple closed curve in 2D. The blue curve segment constitutes its pseudo-concave subset. Image courtesy  ~\cite{doi:10.1111/cgf.13589}.
}
	\label{fig:pseudo}
\end{figure}

Let $E$ be the set of all open connected regions of $Convex hull(\Sigma)\setminus \Sigma$. Each region given by the closure $E$ is defined as a pseudo-concave region (PCR) of $\Sigma$ (Figure \ref{fig:pseudo}). The subset of $\Sigma$ in each PCR is called a \emph{pseudo-concavity}. The edges of the convex hull of $\Sigma$ in each PCR are called convex hull bi-tangents. Based on the radii of medial balls, Peethambaran et al.~\cite{PEETHAMBARAN201562} define \emph{divergent pseudo-concavity} for simple closed planar curves. A pseudo-concave subset of a curve $\Sigma$ is \emph{divergent} if the radii of medial balls monotonically increase as they go along the outer medial axis from one end to the convex hull bi-tangents' end. The curve $\Sigma$ is said to be divergent if all its pseudo-concave subsets are divergent.

From Section 4  in this paper~\cite{ohrhallinger2016hnn} we repeat the following definitions:

The {\em medial axis} $M$ of $\Sigma$ is the closure of all points in $\mathbb{R}^2$ with two or more closest points in $\Sigma$~\cite{blum67medialaxis}. A \emph{medial ball} $B(c, r)$, centered at $c \in M$ of $\Sigma$ with radius $r$, is a maximal ball whose interior contains no points of $\Sigma$.

A smooth curve $C$ (as opposed to $\Sigma$ which may contain inflection points and sharp corners) is a (collection of) bounded 1-manifold(s) embedded in $\mathbb{R}^2$, which are twice-differentiable everywhere except perhaps at boundaries~\cite{duarte2014smoothness}.

We define the {\em local feature size} $\mbox{lfs}(p)$ for a point $p \in C$ as the Euclidean distance from $p$ to its closest point $m$ of $M$. This definition is loosely based on~\cite{ruppert93lfs}, but simplified because we are only considering smooth curves.

Definition~\ref{definition:condition-amenta} is a widely used sampling condition~\cite{amenta98curve} that captures features regardless of size as opposed to globally uniform sampling distances:
\begin{definition}
\label{definition:condition-amenta}
A smooth curve $C$ is \emph{$\epsilon$-sampled} by point set $S$ if every point $p \in C$ is closer to a sample than an $\epsilon$-fraction of its local feature size:
$\forall p \in C, \exists s \in S : \|p, s\|<\epsilon \, \lfs(p)$.
\end{definition}

In contrast, the \emph{reach}~\cite{federer59curvature} for a set $\mathcal S$ is the largest ``radius'' $r$ such that points closer than $r$ to $\mathcal S$ have a unique closest point of $\mathcal S$. The reach is similar to the smallest distance to the medial axis.
This inspires our definition of the \emph{reach} of a curve interval $I$ as $\inf \lfs(p) : p \in I$, where the $\lfs$ is defined by all of $C$.~\cite{ohrhallinger2016hnn}



\subsection{Proximity Graphs}
In general, proximity graphs such as the relative neighborhood graph (RNG), Gabriel graph, Sphere-of-Influence graph \cite{Toussaint} and $\beta$-skeletons \cite{kirkpatrick85beta} play a vital role in defining the shape and structure of planar point sets including curve samples\cite{RNG}. Since many reconstruction algorithms and accompanying theory \cite{edelsbrunner1994three, veltkamp1992gamma,figueiredo94curve, boissonat1984representing} are built around proximity graphs, it seem appropriate to formally define these proximity structures for a better review and understanding of various reconstruction algorithms.

The \emph{relative neighborhood graph} consists of edges $(p, q)$ such that $d(p, q)\leq d(p, x)$ and $d(p, q)\leq d(q, x)$ $\forall$ $x\in P$ where $x\neq p$ or $q$ \cite{RNG}. Varying the size ($\beta$) of region of influence of each pair of points in RNG generates a set of neighborhood graphs called $\beta$-\emph{skeletons} \cite{kirkpatrick85beta}. In $\beta$-skeletons, the neighborhood $U_{p, q}(\beta)$ of two points $p$ and $q$ for any fixed $\beta$ ($0\leq \beta \leq \infty$) is defined as the intersection of two spheres \cite{RNG} as follows: \\
$U_{p, q}(\beta) = B((1-\frac{\beta}{2})p+\frac{\beta}{2}q, \frac{\beta}{2}d(p, q))\cap B((1-\frac{\beta}{2})q+\frac{\beta}{2}p, \frac{\beta}{2}d(p, q)).$

A \emph{minimal spanning tree} MST(P) of $P$ is a tree (a cycle-free graph) that spans all the points in $P$ with the least total cost of edge weights. The \emph{Gabriel graph} of $P$ contains an edge $(p, q)$ if and only if the ball passing through $p$ and $q$ centered at the edge $(p, q)$ is empty \cite{RNG}. A triangulation of $P$ is a subdivision of the plane by edges between vertices in $P$ such that no edge connecting two vertices in $P$ can be added in the plane without creating a self-intersection.
\begin{figure}[h]
	\centering
	\includegraphics[width=7cm]{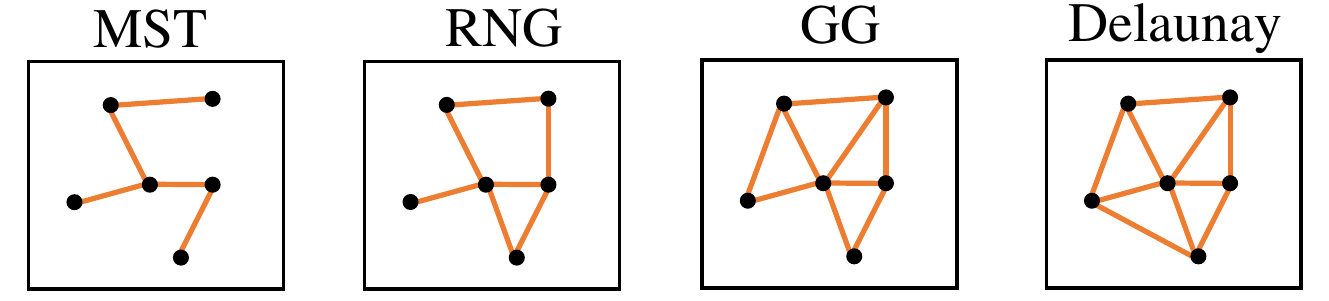}
	\caption{Examples of proximity graphs for a point set.}
	\label{fig:prox}
\end{figure}

A Delaunay triangulation is a triangulation $D(P)$ such that no point in $P$ is inside the circumcircle of any triangle in $D(P)$. A graph is a \emph{Delaunay graph} if it is the Delaunay triangulation of some set of points in the plane. The Delaunay graph has many interesting properties that make it a central data structure in many curve reconstruction algorithms.  The relationship among different proximity graphs, i.e., $MST(P)\subseteq RNG(P)\subseteq GG(P)\subseteq D(P)$ (see Figure \ref{fig:prox}) is a well established result \cite{RNG}. 

For $p\in P$, let $r_{p}$ be the minimum distance from $p$ to $P\setminus p$, and let $B(p, r_{p})$ denotes the open ball of radius $r_{p}$ centered at $p$. Then there exists an edge $(p, q)$ in the \emph{sphere of influence graph} of $P$ if and only if $d(p, q)\leq r_{p}+r_{q}$ \cite{Toussaint}.  These definitions may result in a graph rather than a simple polygon and may contain disconnected regions, non-manifold edges and vertices.
\section{Explicit Reconstruction of Curves}
\label{sec:algorithms}


In this section, we describe the algorithms, grouped by categories.
The first set of reconstruction algorithms (see Subsection~\ref{subsec:graph-based}) were developed {\em based on the proximity graphs} mentioned above.
Then, the introduction of a {\em feature size based sampling condition} permitted to reconstruct features of arbitrary size (see Subsection~\ref{subsec:feature-size}).
These algorithms were then extended to handle {\em noisy samples}  (see Subsection~\ref{subsec:noisy-fitting}) and {\em sharp corners} (see Subsection~\ref{subsec:sharp-features}).
The {\em traveling salesman problem} (see Subsection~\ref{subsec:tsp}) solves a special case of curve reconstruction.
Some algorithm can even reconstruct {\em self-intersecting curves} (see Subsection~\ref{subsec:intersecting}) which allows for new applications.
The Gestalt laws of perception led to {\em curve generation based on the Human Visual System} (see Subsection~\ref{subsec:hvs-based}).
Finally, curves can be fitted approximately to an {\em implicit function} (see Section~\ref{sec:implicit-fitting}) describing the underlying curve.

\subsection{Graph-Based Reconstruction}
\label{subsec:graph-based}

Jarvis~\cite{jarvis1977computing} was the first to develop a notion of shape for a dense unorganized point set in a plane.

This was later formalized by Edelsbrunner et al.~\cite{edelsbrunner83alpha}.
They defined $\alpha$-shapes as a generalization of the {\em convex hull} which permits replacing edges shorter than a globally uniform constant with the opposite two edges of the containing triangle in the Delaunay triangulation of the points.

Edelsbrunner and M\"ucke~\cite{edelsbrunner1994three} later extended this concept to $\mathbb{R}^3$.
It can intuitively be understood as a mass of soft ice cream (the convex hull) containing hard chocolate chips (the points) from which an ball-shaped spoon with radius $\alpha$ nibbles off the ice cream where it can move freely between the chocolate chips, leaving the $\alpha$-shape (which is equal to the convex hull for the case $alpha=\infty$).

Bernardini and Bajaj~\cite{bernardini1997sampling} used that definition to design a construction algorithm for $\alpha$-shapes in $\mathbb{R}^2$: It rotates a disk of radius $\sqrt(\alpha)$ around a point until it touches another point which is then connected by an edge, and continues with the new point, until a loop is created.

Later Bernardini et al.~\cite{bernardini1999ball} developed an extension to $\mathbb{R}^3$, the {\em ball-pivoting} algorithm.

For point sets which are sampled on a curve (as opposed to dense sampling inside the shape as well), edges reconstructing this curve can be selected using the definition of the $\beta$-skeleton~\cite{kirkpatrick85beta}:
All edges of the Delaunay triangulation of the point set that are shorter than $\frac{\beta}{2}$ times the radii of the circumcircles of their adjacent triangles belong to this $\beta$-skeleton.

Veltkamp proposed the $\gamma$-neighborhood graph~\cite{veltkamp1992gamma} which unifies the convex hull, the Delaunay triangulation, the {\em Gabriel graph} and the $\beta$-skeleton, and presents the relations to and between other neighborhood graphs.

Based on the $\gamma$-neighborhood graph, he later showed that a greedy algorithm based on both local and global measures can also reconstruct boundary polygons which are not in the Delaunay triangulation~\cite{veltkamp19933d}.

This was necessary because Delaunay triangulations do not always contain a {\em Hamiltonian cycle}, a simple polygon interpolating all its points, although that case has been shown to be very rare~\cite{genoud90hamiltonian}.

Boissonat~\cite{boissonat1984representing} used {\em sculpturing} to replace edges in the convex hull with their Delaunay triangle counterparts (same as for $\alpha$-shapes), successively ordered by diminishing edge length.
While this still guarantees the resulting polygon to remain manifold, it can only reconstruct a single curve.
The algorithm may also get stuck in case densely sampled points remain in its interior, instead of being interpolated.

O'Rourke~\cite{o1987connect} proposed to compute the minimal length tree in the Voronoi diagram corresponding to a polygonal boundary in its Delaunay triangulation dual, but this requires a distinct skeleton to work.

De Figueiredo and Gomes~\cite{figueiredo94curve} proved that the {\em Euclidean minimal spanning tree} (EMST) reconstructs open curves from sufficiently dense samples, with density defined by an empty tubular neighborhood.

Attali defined an {\em r}-regular shape~\cite{attali97regular} as having a boundary with curvature $\ge r$ everywhere.
Then they proved its reconstruction from a uniform sampling of that boundary such that all discs with radius $<\frac{r}{2}$ centered on the boundary contain at least one sample.

Stelldinger~\cite{stelldinger2008topologically} proved both the correctness of the above-mentioned ball-pivoting algorithm~\cite{bernardini1999ball} and a much more strict bound on required sampling of the points, but only for the uniformly sampled case: minimum 6 points for a sphere, as opposed to 22~\cite{niyogi2008finding}, 484 ($\epsilon=0.1$,~\cite{amenta1998new}) or 1343 ($\epsilon=0.06$,~\cite{amenta2000simple}).

Stelldinger and Tcherniavski~\cite{stelldinger2009provably} extended the proof above to noisy uniform samples.

Ohrhallinger and Mudur~\cite{ohrhallinger11operations} also exploited the minimum length property of the EMST.
They used edge exchange operations to transform it into a manifold.
While they show successful reconstruction for several sparsely sampled and noisy point sets, there is no easily applicable sampling condition.
Its time complexity is non-polynomial in principle even if for many cases it terminates in $O(N log N)$ time.

In a later paper~\cite{ohrhallinger13connect2d}, they define a modification of the EMST which relaxes its vertex valence constraint from $\ge 1$ to $\ge 2$.
This {\em minimum boundary complex} can be approximated well in $O(n log n)$ time.
By inflating (a dual to the  sculpturing~\cite{boissonat1984representing} operation) they achieve a manifold boundary already close to the points.
This facilitates the subsequent sculpturing step, with much reduced risk of falling into local minima.
They proved correct reconstruction for a tightened sampling condition of $\epsilon<\frac{1}{2}$, although it additionally requires a local uniformity $u<1.609$ as expressed in the proportion of the lengths of adjacent edges.

Peethambaran et al.~\cite{PEETHAMBARAN201562} defined a proximity graph called \emph{shape hull graph (SHG)} which faithfully reconstructs smooth curves that exhibit divergent concavity. The authors characterize the divergent concave curves based on the exterior medial balls in the pseudo-concave~\cite{PEETHAMBARAN201562} areas. The algorithm constructs the SHG by repeatedly removing boundary Delaunay edges subjected to geometric and topological properties. The geometric criterion (circumcenter location of the Delaunay triangle) is used to prune off elongated Delaunay triangles whose vertices lie further apart from each other on the curve and the regularity criterion eliminates non-manifold elements, e.g., dangling edges and junction points, in the resultant polygon, thereby making it topologically equivalent to a circle (sphere in 3D).

In a subsequent work, the authors~\cite{doi:10.1111/cgf.13589} employed an incremental algorithm to classify Voronoi vertices into \emph{inner} and \emph{outer} with the help of normals estimated through Voronoi poles. Such a classification not only helps reconstructing the underlying curve, but also aids in medial axis computation and dominant point detection. Theoretical guarantees under \emph{bi-tangent neighborhood convergence}, a slightly modified version of divergent concavity for simple closed and planar curves, is also provided.

In a greedy approach introduced by Parakkat and Muthuganapathy~\cite{parakkat2016crawl}, starting from the smallest edge in the Delaunay triangulation (which is guaranteed to be part of the reconstructed curve under $\epsilon$-sampling), the algorithm iteratively adds an appropriate shortest edge to the result until it satisfies some conditions. The procedure is repeated to facilitate the capturing of disconnected components. Also, they employ a heuristic to identify whether the reconstructed curve is open or not.

Graph-based curve reconstruction methods often require the user to choose a global parameter, and in consequence they yield good results only for uniformly sampled points.
This means that for samples spaced too widely apart, the reconstructed curve may contain holes.
On the other hand, if the spacing is too dense, not all samples may be interpolated by the output.
On top of that, there is no guarantee that a Delaunay graph contains a polygon interpolating all samples.

\subsection{Feature Size Criteria Reconstruction}
\label{subsec:feature-size}

Amenta et al. proposed in their seminal paper~\cite{amenta98curve} to apply the concept of the {\em local feature size}~\cite{ruppert93lfs} to the spacing of samples and define the {\em Crust} as a subset of the Delaunay triangulation of the point set.
The {\scshape Crust} algorithm which reconstructs this curve does not require the user to tune a global parameter for the (uniform) sample spacing.
Instead it permits reconstruction of this subset from non-uniformly sampled points which is a curve as long as they conform to their stated sampling condition.
This sampling condition requires a minimum angle between adjacent edges (assuming equal edge lengths) of the reconstructed piece-wise curve, which increases with their proportion of lengths.
{\scshape Crust} requires an $\epsilon$-sampling of $\epsilon<0.252$ which corresponds to an angle $\alpha > 151.05^\circ$ between adjacent edges.
While it is important as a theoretical result, in practice these angle requirements are quite restrictive and difficult to ensure for point sets.

Gold~\cite{gold99anticrust} developed a one-step algorithm which extracts above {\em Crust} without having to construct the Voronoi diagram on top of the Delaunay triangulation, and with it, the {\em Anti-Crust}, the skeleton approximating the medial axis.

Dey and Kumar~\cite{dey99curve} improved on that result with the elegant and simple {\scshape NN-Crust} algorithm which relaxes the sampling condition to $\epsilon<\frac{1}{3}$, corresponding to $\alpha>141.62^\circ$.
It first connects the points to their nearest neighbors, and then adds a second edge, where necessary, to the nearest point such that it creates an angle $>90^\circ$.

Dey et al.~\cite{dey99conservative} extended {\scshape Crust} as well, to {\scshape Conservative Crust}, which filters specific edges from the Gabriel graph.
It is able to reconstruct (collections of closed and) open curves, and it is also robust to outliers.
However, it requires a parameter and misses some sharp corners which can be reconstructed by {\scshape Crust} and {\scshape NN-Crust}.

Lenz~\cite{lenz06curve} claimes to relax the required density of {\scshape NN-Crust} to $\epsilon<0.4$ and up to $\epsilon<0.48$ depending on the angle $\alpha$ but without proof.
The proposed algorithm also permits the reconstruction of sharp corners and self-intersecting curves, starting with a seed edge between the two closest points and connecting edges by tracing along with a probe shape.

Hiyoshi~\cite{hiyoshi09optimize} adapted the Traveling Salesman Problem to multiply connected curves, thus making it solvable in polynomial time as a maximum-weight 2-factor problem. The algorithm operates on the Delaunay triangulation and proved correct reconstruction for $\epsilon<\frac{1}{3}$ and relative uniformity of adjacent edge lengths, differing at most by a factor of 1.4656.

Ohrhallinger et al.~\cite{ohrhallinger2016hnn} described a simple variant of {\scshape NN-Crust} which they call {\scshape HNN-Crust} since they connect both nearest neighbor and a so-called {\em half neighbor} per point (unless it forms an end point of the curve).
This half neighbor is defined as the nearest point lying in the half space opposite the bisecting edge of the nearest neighbor edge.
Connecting these neighbor points reduces the minimum angle from $90^\circ$ (for {\scshape NN-Crust}) to $60^\circ$.
They improve the sampling condition for this algorithm up to $\epsilon<0.47$. Furthermore, they introduce a new reach-based sampling condition which they relate to $\epsilon$-sampling, $\rho=\frac{\epsilon}{1-\epsilon}$. It manages to reduce the number of required samples for reconstruction and to permit sharp angles by defining the distance to the medial axis at intervals between samples instead of at samples only.

\subsection{Fitting Curves to Noisy Points}
\label{subsec:noisy-fitting}

The above-mentioned algorithms do not reconstruct curves well if the samples are contaminated by noise.

Lee~\cite{lee2000curve} uses a technique called {\em moving least-squares (MLS)}~\cite{levin1998approximation} which iteratively projects points on a curve fitting their local neighborhood by distance-weighted regression.
This results in a thinned point cloud which can be locally approximated by a line inside a constant-sized neighborhood so that the center can be connected with its furthest neighbor points to form the edges of the reconstruction.
The weighting function for the MLS projection considers points only inside a globally constant radius, which could also include unwanted points.
Therefore the connectivity of points is created using the EMST, which minimizes edge length, and is then traversed to determine the local noise extent.

An noise-robust extension~\cite{mehra2010visibility} of the {\em Hidden Point Removal (HPR)} operator~\cite{katz2007direct} computes local connectivity between points based on a projection onto their convex hull.
The global reconstruction is then extracted by approximating the maximum weight cycle from a weighted graph combining the local connectivity.
The algorithm does not denoise or smoothen, i.e. simply interpolates points, and the reconstruction exhibits holes or misses points in regions with moderate noise extent.

Rupniewski~\cite{rupniewski2014curve} first sub-samples a noisy point set with minimum (globally constant) density.
Then he uses a heuristic that alternatingly moves these points to local centers of mass based on the Voronoi diagram and eliminates points which do not have exactly two neighbors in a density-sized neighborhood, until the point set is stable. Finally, after hundreds of iterations, the points can be consecutively ordered. Only very basic results are shown in the paper.

Cheng et al.~\cite{cheng2005curve} resample a thinned point set from noisy points and then use {\scshape NN-Crust} to reconstruct the curve. They prove a probabilistic sampling condition, however it is impractical due to its restrictive sampling density constraints, and they only prove but do not show any results of their proposed algorith

Wang et al.~\cite{wang2014robust} first construct a quad-tree on the samples to determine inner and outer boundaries of noisy samples on a grid.
After smoothing these boundaries, they compute their Voronoi diagrams in order to extract the skeleton which represents the reconstructed curve. While their method is very resilient to outliers and noise, it requires careful tuning of several parameters and does not handle sparse samples well.

{\scshape FitConnect}~\cite{ohrhallinger2018fitconnect} seamlessly extends parameter-free {\scshape HNN-Crust} to handling noisy samples.
The conforming condition of the latter which specifies whether three points can be connected in exactly a single way is extended by fitting a circular arc to the local neighborhood if consisting of more than three points.
Where local fits do not overlap consistently, they are grown to larger neighborhoods until covering these noisy clusters.
The resulting ordered consistent local fits are then denoised to this locally estimated noise extent (the variance of the fits) by blending them together.
The algorithm also manages to classify sharp corners which would otherwise be smoothed.
Its runtime is however O($k^2$) in the size $k$ of noisy neighborhoods.

{\scshape StretchDenoise}~\cite{ohrhallinger2018stretchdenoise} improves the blending technique used for denoising in {\scshape FitConnect} by modeling the recovered manifold connectivity separated from the high-frequency residuals.
These are used to shift point positions by minimizing angles between edges in the least-squares sense.
Additionally, movement of points is restricted to lie inside a probability density function cut-off distance, which is estimated from the variance of the fitted arcs but can also be input from sensor noise models.
This also guarantees stochastic error bounds for the noisy samples.

Fitting curves to approximate noisy samples is a difficult task and trades off recovering feature detail vs. robustness.

It is worth noting that, for some applications, instead of a polygonal reconstruction, they prefer the reconstructed result to be a polynomial curve(s). A few among such applications include vector representation of computer fonts \cite{nonpoly1} and reconstructing the contours of medical images \cite{nonpoly2}. Unlike polygonal reconstruction methods (which is the main focus of this report - and uses straight lines to connect appropriate points), these methods fit the input points by polynomial curves (mainly B-Splines or Bezier curves) that minimize a particular cost function.


\subsection{Reconstructing Sharp Features}
\label{subsec:sharp-features}

The well-known and established $\epsilon$-sampling condition has a significant drawback; it cannot sample a sharp corner. That is because the medial axis touches the corner and hence would require an infinite amount of samples at that particular point to satisfy $\epsilon$-sampling for any $\epsilon$.

Assuming a new sampling condition based on the tangential circles with respect to a point in the curve (to avoid the need of infinite sampling at sharp corners), {\scshape Gathan}~\cite{dey01corners} modifies the nearest neighbor strategy to handle sharp corners. While selecting an edge $e$, the improved algorithm takes into account the angle between the dual Voronoi edge and estimated normal of $e$, ratio of its dual Voronoi edge length to its length and the degree of all the vertices. This method is later on extended~\cite{dey02gathang} by carefully structuring it to provide a theoretical guarantee. The improved algorithm requires only one parameter, which gives the minimum angle of all sharp corners based on which the sharp corners are locally sampled.

Rather than imposing extra conditions, Funke et al. ~\cite{funke01curve} proposed an algorithm which is guaranteed to reconstruct the curve faithfully under a specific sampling condition. They proposed a sampling condition relying on the edges of the correct reconstruction for a smooth curve and later on relax it at corner points to generate a weak sampling around it. Starting from justifiably `smooth' edges, their reconstruction explores potential corners. The identified corner edges are then merged with the smooth edges to give the final reconstructed result.

While algorithms specialized to handle sharp features reconstruct these cases quite well, their conditions are often complex, and they do not compete well for the general case.


\subsection{Traveling Salesman Methods}
\label{subsec:tsp}


Giesen~\cite{giesen99delaunay} showed in an existence guarantee that for sufficiently dense sampling the boundary can be reconstructed by solving the {\em Euclidean Traveling Salesman Problem} (ETSP) for a set of points.
He proposes two algorithms in that paper, but does not present any results.

Althaus et al.~\cite{althaus00polynomial} add to this that it also works for non-uniform sampling.
Furthermore, they show that if constrained by an $\epsilon$-sampling~\cite{amenta98curve}, the NP-hard ETSP terminates in polynomial time.
However, they manage to prove that only for a very restrictive $\epsilon<\frac{1}{20}$, which permits just angles $>174.27^\circ$.

Althaus et al.~\cite{althaus00curve} compared approximation algorithms for the ETSP, based on heuristics, but noted that these all fail for sparsely sampled cases where the ETSP would have succeeded.
An interesting observation is that the complexity for the ETSP construction decreases as the sampling gets more dense.
A naive ETSP construction takes $O(2^n)$ time which is unfeasible even for very small point sets.

Arora et al.~\cite{arora1998polynomial} approximate the ETSP within $(1+\frac{1}{c})$ in $O(n (log n)^{O(c)})$ time, but their reconstructions result in poor visual quality.
The fastest exact TSP solver, the {\em Concorde}~\cite{url:concorde}, would still take years to compute the boundary for practical point sets, as can be derived from a discussion of its complexity~\cite{hoos2009empirical}.


In analogy to the Traveling Salesman problem of minimizing curve length, polyhedra with minimal area were proposed for surface reconstruction~\cite{o1981polyhedra}. But for this NP-hard problem no algorithm exists, furthermore Boissonat showed by a simple example that this minimum is not always visually pleasing~\cite{boissonnat1984geometric}.

The runtime performance of the ETSP degrades drastically for sparsely sampled points.
Those point configurations tend to be rather ambiguous w.r.t. which points are connected to the boundary.
The ETSP is therefore not a suitable tool for curve reconstruction as it puts almost all computational effort where it makes little difference in terms of visual aesthetics.

\subsection{Curves with Self-Intersections}
\label{subsec:intersecting}

Most of the curve reconstruction work concentrates on reconstructing a (set of) simple closed and/or open curve(s). But for applications like sketching~\cite{parakkat2018peeling} or point sets generated from images, the inputs might also contain self-intersections.

The first one in this category~\cite{degoes2011optimal} formulates and solves an optimal transport problem. Starting from the Delaunay triangulation of the input point set, their approach generates a coarse mesh in a greedy fashion with the objective of minimizing the total cost. With the help of intelligent vertex relocation, this approach is specially designed to handle noise and outliers. Since this procedure does not impose any manifold or degree constraints on the input while filtering out edges from the simplified mesh, it can reconstruct shapes with self-intersections.

Instead of modifying the algorithm itself to adapt to reconstructing curves with self-intersections, Parakkat et al.~\cite{parakkat2018peeling} use a post-processing step to identify and restore self-intersections. Initially, a reconstruction step with a vertex degree constraint of maximum three is used. Later on, potential intersections are explored at the vertices with degree one. Based on a user parameter and the one-ring Delaunay neighborhood of the considered vertex, potential self-intersections are recovered by the appropriate Delaunay edges.

It is worth mentioning that even if {\scshape Crust}~\cite{amenta98curve} and {\scshape NN-Crust}~\cite{dey99curve} are not particularly designed to handle curves with intersections, in some cases they capture self-intersections since they do not impose a manifold restriction on the vertex degree.


\subsection{Curve Generation based on Human Visual System}
\label{subsec:hvs-based}

A few curve reconstruction algorithms rely on a subset of Gestalt laws of perception which describe how humans perceive visual elements. Among the six Gestalt rules, \emph{proximity} and \emph{continuation} are very important to curve reconstruction strategies. While the proximity rule suggests that the human visual system has a natural tendency to group nearest points, the law of continuation states that the human eyes will follow the smoothest path when viewing curves and hence helps guiding our eyes in a certain direction while connecting the points~\cite{mather2016foundations}. Following the Gestalt law of proximity, Zeng et al.~\cite{zeng08distance} proposed a parameter free algorithm called {\scshape DISCUR} for reconstructing multiple simple curves that may be closed or open as well as contain sharp features. A successful reconstruction using {\scshape DISCUR} depends on an appropriate sampling of interior curves and an accurate identification of boundary curves. Since {\scshape DISCUR} relies on the proximity criterion, wrong connections may occur when a sample has two or more nearest neighbors. In such cases, the selection is quite arbitrary. Furthermore, it requires a very dense sampling near sharp corners in order to correctly reconstruct these.

An improved version of {\scshape DISCUR} has been presented as well~\cite{nguyen08vicur}. The authors utilize the Gestalt principles of proximity and continuity to formulate a vision function that is supposed to best mimic the natural human vision. The main intuition behind the algorithm is that any abrupt changes while connecting the points are reflected in the statistical properties of the curve, which are in turn captured through the vision function. The algorithm, known as {\scshape VICUR}, employs appropriate rules based on the vision function to reconstruct multiple closed or open curves with or without sharp features. A drawback of VICUR algorithm is that it is highly sensitive to the user-tuned parameters.

Note that Edge exchanging~\cite{ohrhallinger11operations} and Connect2D~\cite{ohrhallinger13connect2d} algorithms (already described in Subsection~\ref{subsec:graph-based} both also relate to the Gestalt laws.


\section{Implicit Curve Fitting}
\label{sec:implicit-fitting}

Implicit functions for curve or surface fitting has been widely investigated in the computer graphics community. Implicit methods attempts to define a smooth function $f \colon$ $\mathds{R}$$^{2}\rightarrow$ $\mathds{R}$ such that the zero level set of $f$ approximates the underlying curve in the input points as illustrated in Figure \ref{fig:implicit}. The zero level set of the curve (also referred to as contour) or surface can be directly visualized by using a ray tracer or by polygonizing it using the well-known marching squares or cubes algorithm~\cite{10.1145/37402.37422}. Many algorithms have been developed for algebraic curve fitting to a set of 2D or 3D points, e.g., conic planar curves ~\cite{BOOKSTEIN197956, FORSYTH1991130} and curves of arbitrary degrees~\cite{Vaughan87, Taubin, Taubin1}. A curve or surface is \emph{algebraic} if their representative functions, i.e., $f$ are polynomials of some degree $d$. Fitting algebraic curves to a finite set of points is normally posed as a least square fitting problem where the objective is to minimize the mean square distance from the sample points to the curve.

In general, implicit functions are extremely compact and suitable for representing free-form curves \cite{10.1145/2732197}. Though most of the implicit techniques focus on surface fitting, many of them can be either directly applied or adapted for planar curve fitting. A typical choice for the implicit function is the \emph{Signed distance function (SDF)}. The SDF for an arbitrary point $p$ is the signed distance between $p$ and its nearest point on the boundary where the sign component indicates the location of $p$ with respect to the curve, i.e, whether the point lies inside or outside the boundary. Reconstruction methods employing SDF range from tangent plane estimation \cite{Hoppe:1992:SRU:142920.134011} to polynomial splines over hierarchical T-meshes \cite{5521470}.

Radial basis functions (RBF) represent an excellent tool for the smooth interpolation of scattered points.  Carr et al.~\cite{Carr:2001:RRO:383259.383266} formulate the implicit function $f$ as a linear sum of weighted and shifted radial functions, i.e., $f(p)=\sum_{i=1}^{n}w_{i}\phi(\|p-c_{i}\|)$, where the weights $w_{i}$ are determined by solving a linear system constructed from various surface constraints at input points $c_{i}$. Choices for the basic function $\phi$ include Gaussian ($\phi(r)=exp(-cr^{2})$), multi-quadric ($\phi(r)=\sqrt{c^{2}+r^{2}}$), polyharmonic ($\phi(r)=r$ or $\phi(r)=r^{2}$) and thin-plate spline ($\phi(r)=r^{2}\log r$). In a related work~\cite{Turk:2002:MIS:571647.571650}, the authors estimate the implicit surface as a RBF that minimizes thin plate energy subject to a set of interior and exterior constraints.

Poisson reconstruction~\cite{10.5555/1281957.1281965} solves for an indicator function for the curve (or surface), whose gradient best approximates the normal field $N$, i.e., $F=$argmin$_{S}\parallel\nabla_{S}-N \parallel_{2}^{2}$. This optimization problem leads to a Poisson equation, which is solved by a locally supported radial basis function on an adaptive octree (quadtree in 2D). As the current gold standard in the community for surface reconstruction, it however requires normals at points to be specified. In related works, Fourier~\cite{10.5555/1281920.1281931} and wavelet~\cite{doi:10.1111/j.1467-8659.2008.01281.x} bases have been employed for an accelerated solving of Poisson equations.

An alternative to RBF curve interpolation is the moving least square (MLS) method. The MLS projection \cite{Levin2004} method first defines a local reference frame $\mathcal{H}$ for a point $q$ to be projected and then fits a local polynomial approximation $g$ to the weighted input points. Here, the weight for each input sample $p_{i}$ is a function of its distance to the projected $q$ on $\mathcal{H}$. Once the polynomial is computed, the projection of $q$ onto $g$ represents the MLS projection of $q$. Above-mentioned Lee ~\cite{LEE2000161} describes an improved moving least-squares technique using Euclidean minimum spanning tree and region expansion for fitting non-intersecting curves to unorganized point clouds. Several MLS based approaches including surface re-sampling \cite{Alexa03computingand}, progressive point set surfaces \cite{Fleishman:2003:PPS:944020.944023}, sharp feature reconstruction \cite{Fleishman:2005:RML:1073204.1073227}, algebraic spheres (or circles) \cite{Guennebaud:2007:APS:1276377.1276406},  provable MLS surfaces \cite{Kolluri:2008:PGM:1361192.1361195}, have been proposed. Being insensitive to noise, MLS approaches are suitable for fitting curves to noisy data.

The idea of decomposing the input data domain into sub-domains and locally fitting piece-wise quadratic functions to the data is prevalent in the surface reconstruction and is equally applicable to 2D curve fitting. Ohtake et al.~\cite{Ohtake:2003:MPU:882262.882293,Ohtake200615} blend the locally fitted quadratic functions using a weighing function (the partitions of unity) to create the global approximation to the underlying surface. Alliez et al.~\cite{alliez2007voronoi} utilize a Voronoi diagram of the input point set to deduce a tensor field whose principal axes and eccentricities locally represent respectively the most likely direction of the normal to the surface and the confidence in this direction estimation. An implicit function is then computed by solving a generalized eigenvalue problem such that its gradient is most aligned with the principal axes of the tensor field, providing a best-fitting iso-surface or curve reconstruction.

In general, implicit curves employ acquired or estimated point normals to facilitate the reconstruction process, and are found to be robust against noise. However, since the iso-curves are extracted using marching squares on quadtrees, an appropriate quadtree depth has to be determined and preset in the case of implicit methods. Detailed curves can be generated for larger depth values, however at the expense of increased computational time. While popular in surface reconstruction, we have not found implementations or results for curve reconstruction, and so this category is excluded from our comparison or evaluation.

\section{The Benchmark}
In this section, we briefly describe the curve reconstruction benchmark. The motivation behind setting up such a benchmark is to encourage the use of standardized datasets and evaluation criteria for research and advancements in curve reconstruction and related applications. The proposed benchmark repository consists of a driver program, data sets, associated ground truth, sampling and evaluation tools in the shape of test scripts for a comprehensive experiment on 2D curve reconstruction algorithms. We also provide a set of publicly available curve reconstruction algorithms in the benchmark. The selected algorithms include early ones from the late nineties up to recent papers. The components of the benchmark repository and their interactions are illustrated in Figure \ref{fig:teaser} and discussed in the following sections (Sections \ref{sec:b_algo}-\ref{sec:b_script}).

\subsection{Algorithms}\label{sec:b_algo}
We have included a set of fifteen publicly available curve reconstruction algorithms in the benchmark. Table \ref{table:compared-algorithms} records the list of algorithms and the abbreviations that we use in our experiments to refer to them. Note that we were not able to obtain code for some of the algorithms (~\cite{mehra2010visibility},~\cite{lee2000curve},~\cite{wang2014robust},~\cite{hiyoshi09optimize}) and therefore could not include those in the benchmark. All the algorithms except {\scshape Optimal Transport}~\cite{degoes2011optimal} interpolate or try to closely fit the input points. On the contrary, {\scshape Optimal Transport} focuses on simplified reconstruction and hence, for a fair comparison, we have not included it in any of our experiments, but presented some representative results.
\begin{table}[h]
	\begin{center}
		\begin{tabular}{|l|r|}
			\hline
			Algorithm & Open Source \\
			\hline
			{\scshape crust}~\cite{amenta98curve} & yes \\
			{\scshape nncrust}~\cite{dey99curve} & yes \\
			{\scshape ccrust}~\cite{dey99conservative} & yes \\
			{\scshape gathan}~\cite{dey01corners} & yes \\
			{\scshape gathang}~\cite{dey02gathang} & yes \\
			{\scshape lenz}~\cite{lenz06curve} & yes \\
			{\scshape connect2D}~\cite{ohrhallinger13connect2d} & yes~\cite{url:connect2d} \\
			{\scshape crawl}~\cite{parakkat2016crawl} & yes~\cite{url:crawl} \\
			{\scshape hnncrust}~\cite{ohrhallinger2016hnn} & yes~\cite{url:hnn-crust} \\
			{\scshape fitconnect}~\cite{ohrhallinger2018fitconnect} & yes~\cite{url:fitconnect} \\
			{\scshape stretchdenoise}~\cite{ohrhallinger2018stretchdenoise} & yes~\cite{url:stretchdenoise} \\
			{\scshape peel}~\cite{parakkat2018peeling} & yes~\cite{url:peeling} \\
			{\scshape optimaltransport}~\cite{degoes2011optimal} & yes~\cite{url:optimaltransport} \\
			{\scshape discur}~\cite{zeng08distance} & yes \\
			{\scshape vicur}~\cite{nguyen08vicur} & yes \\
			\hline
		\end{tabular}
		\caption{Algorithms compared in our study. Source is provided together with our benchmark unless referenced here, in which case it will be pulled from the respective repository.}
		\label{table:compared-algorithms}
	\end{center}
\end{table}
\begin{figure}[h]
	\centering
	\includegraphics[width=8cm]{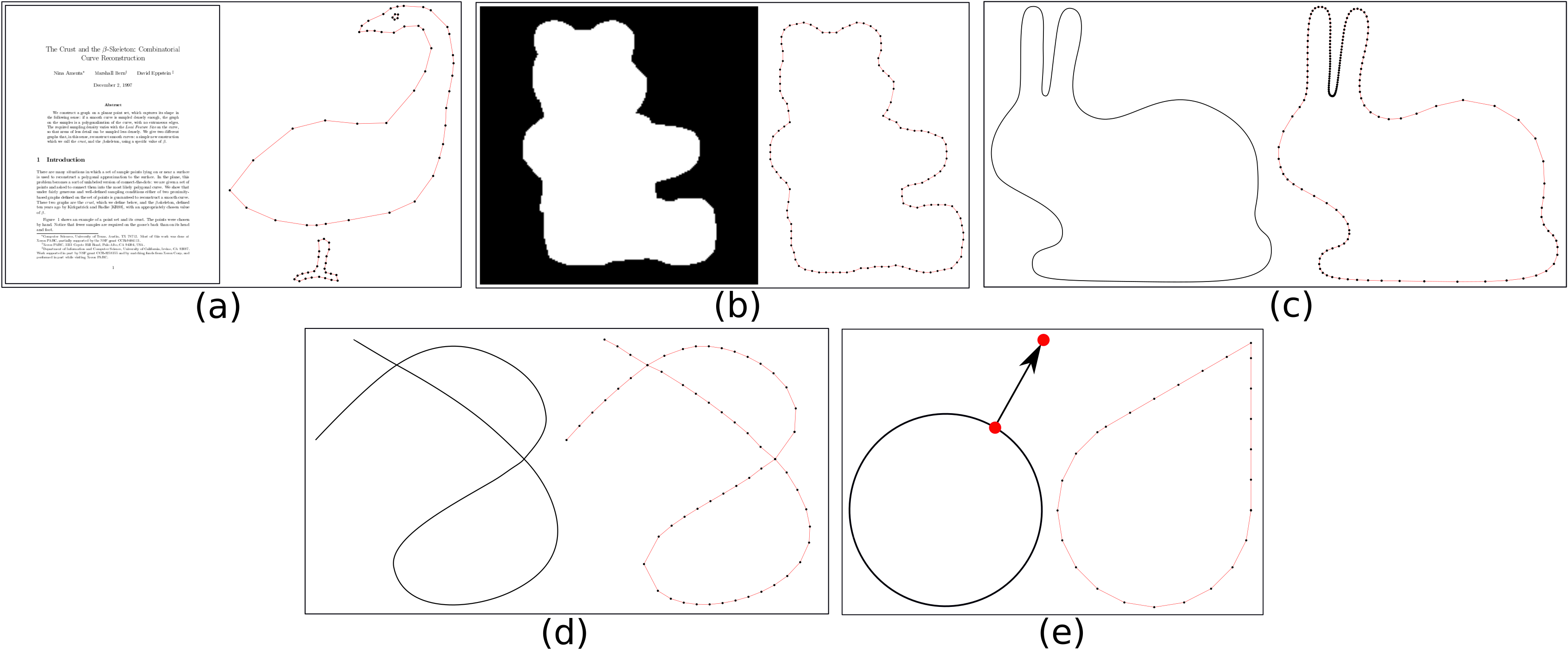}
	\caption{Examples of different types of test data. (a) Classic data collected from different papers, (b) Points sampled from a binary image boundary, (c) LFS-sampling from a cubic B\'ezier curve, (d) Points sampled from a synthetic curve, (e) Synthetic data generated by extruding sharp corners from circles.}
	\label{fig:testdata}
\end{figure}
\subsection{Data Sets and Associated Ground Truth}
\label{sec:b_data}
\label{sec:data}

\begin{figure}[h]
	\centering
	\includegraphics[width=8cm]{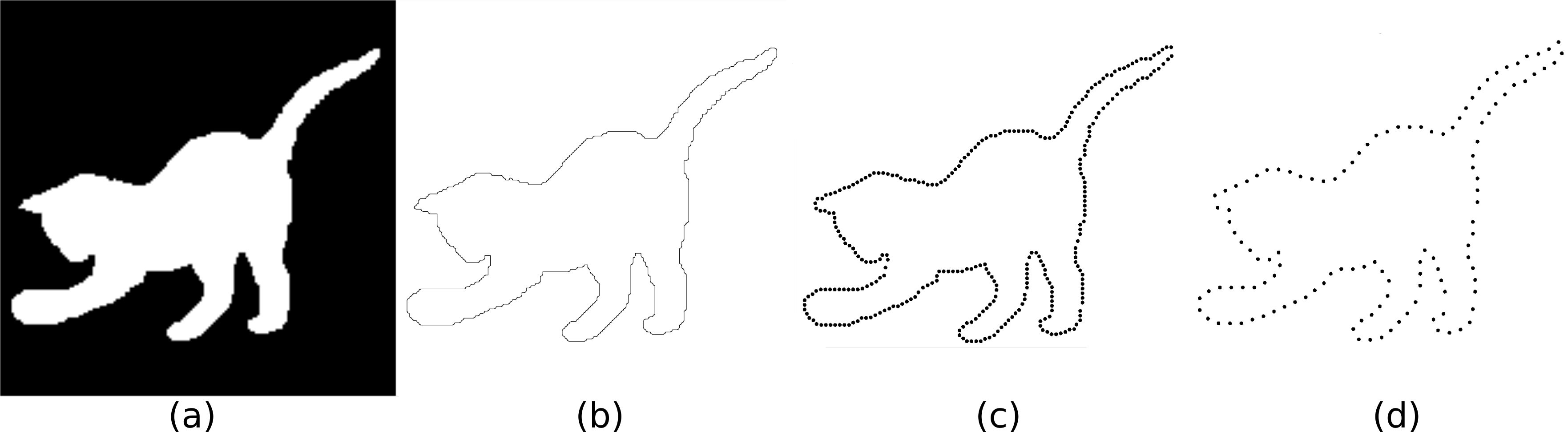}
	\caption{(a) A binary image, (b) Extracted edges, (c) Result of sampling with radius = 20, (d) Same for radius = 50.}
	\label{fig:bdrydata}
\end{figure}

We collected data from various sources as well as synthetically generated test cases using analytical functions. We also provide the ground truth associated with the test data if they represent a linear approximation to the input curve. We classify the test data based on the data source and mode of generation (see Figure \ref{fig:testdata} for examples from the different categories) as follows:

\textbf{\scshape Classic} This set of data consists of all the point sets collected from various curve reconstruction papers and/or projects, mostly from project page or repository. A few data sets were extracted from the images using webplot digitizer~\cite{webplot}. Test data include well known point sets used for evaluating the reconstruction quality of sharp features, open curves and sharp corners. We collected 25 manifold curves, 16 non-manifold curves, 21 curves with sharp corners, 23 open curves and 52 multiply connected curves. 

\textbf{\scshape Image} This set consists of contours extracted from silhouette images (see Figure~\ref{fig:bdrydata}) from various image databases, e.g., MPEG-7 Core Experiment CE-Shape-1 Test Set \cite{mpeg}, Edinburgh Kitchen Utensil Database \cite{etudataset}, and the 1070-Shape Database \cite{1070}. A description of how the contours (2158 manifold, 2 multiply connected, 206 sharp corners) are extracted is given in Subsection \ref{sec:b_tools}.

\textbf{\scshape Synthetic} Two analytical shapes (bunny, sharp corner) were sampled with this method~\cite{ohrhallinger2016hnn}, also detailed in Subsection~\ref{sec:sampling}. These data sets were used for our experiments on feature-sized noise, sampling density and curves with sharp features. The code to generate sharp curves with varying degrees of sharpness (see Figures \ref{fig:testdata}(d)- \ref{fig:testdata}(f)) is available as a part of \emph{CurveBenchmark.cpp}, the driver program.

All the experimental data are organized inside different subdirectories, i.e., {\em multiple-curves, open-curves, sharp-corners, non-manifolds}, and {\em manifolds}. Test data sampled from curves exhibiting multiple features are placed under more than one directory. Each test case that we experimented with has an associated ground truth which is used to evaluate the reconstruction quality. This ground truth is represented using either an \emph{indexed list} or \emph{ordered vertices}. In the indexed list representation, the ground truth file stores all the vertices first followed by the pairs of vertex indices representing the edges. Ordered vertices represent the edges of the curve using consecutive vertices in the file.

\subsection{Sampling Tools}\label{sec:b_tools}
\label{sec:sampling}
In order to analyze how reconstruction algorithms perform w.r.t. varying sampling density, we repeat here a simple sampling algorithm~\cite{ohrhallinger2016hnn} that creates an approximate $\epsilon$-sampling on cubic B{\'e}zier curve input.

First we densely sample the segments of the B{\'e}zier curve along its parametrization.
Then the normal $n_i$ at each curve sample $s_i \in S$ is computed as orthogonal to the edge connecting its neighbor samples on the curve.
The largest empty disc at $s_i$ can be established by $s_i, n_i$ and querying each other curve sample $s_j \in \{S \setminus s_i\}$ by setting the disc center $c_j=s_i+tn_i,\|cs_i\|=\|cs_j\|$. After solving this, the $c_j$ with the largest radius of all empty discs are added to the set of medial axis points $M$.
Now, having sampled this medial axis approximation, we can simply estimate the lfs for each $s_i$ by locating its nearest neighbor in $M$ and its distance.
Note that this computation is not exact due to the discretization of the original curve as well as floating point precision error.
However, computing medial axis and thus the lfs exactly is a hard and computationally expensive task~\cite{aichholzer09medialaxis},~\cite{attali09medialaxis}.
But since $\epsilon$-sampling requires an upper bound on distance, and the curve is also discretized, the chosen samples should be mostly within that bound.
In order to sample the curve with a given $\epsilon$, we now start with any curve sample $s_i$ (or any on its boundary if the curve is open) and iterate over successive samples along the curve while $\|s_i,s_j\|/\mbox{lfs}<\epsilon$ and choose the last valid one as next point in our $\epsilon$-sampling.

Since we have now computed the lfs for all samples, we can further perturb these in relation to the lfs, in order to simulate feature size varying noise.
We retain the sampling density by just moving each sample along their normal which was incidentally determined by the fitting of the empty discs.

We also provide a discrete sampling tool written in Processing3 (\url{www.processing.org/}) for extracting points from a given binary white-on-black image. It first extracts the pixels lying on the object boundary by comparing each pixel with its 8-neighborhood. Then, the extracted boundary is fed to a boundary sampling algorithm. Based on a user given radius $r$ (which determines the sampling density), the sampling algorithm randomly picks a pixel at position (x,y), inserts a point at the location (x,y), and erase all boundary pixels lying at a distance less than $r$ from (x,y). This procedure is repeated until all boundary pixels have been erased. Figure \ref{fig:bdrydata} shows a sample binary image, its extracted edges, and samples generated for two radius values.

\textbf{Interactive sampling:} For some randomly sampled point sets, identifying the ground truth is tricky since no algorithm can claim a proven reconstruction. In such cases, in order to help the user generate the ground truth, we use an interactive ordering program. The interactive ordering program displays the point set, the user selects points by clicking on them, and the points are saved in this user-specified order. This program then also displays the edges according to that order.


\subsection{Evaluation Criteria}\label{sec:b_criteria}
In order to measure how well a reconstructed curve $C'$ approximates the original $C$, we compute the distance for closest points between the two curves, similar to this benchmark~\cite{berger2013benchmark} between both shortest distance maps $M:C \mapsto C'$ and $M':C' \mapsto C$ since the mapping is not bijective: $M' \neq M^{-1}$.

We sample sets of points $S,S'$ on the two respective curves $C,C'$ (which consist of edge-chains) uniformly and densely.
Then for each sample $s \in S$, we determine its closest point $t \in C'$ to create a discrete mapping $(s,t)$ from all $s' \in S'$ to $t' \in C$, and a similar reverse mapping $(s',t')$.

The set of closest point correspondences are then:

\begin{equation}
D={(s,t)|s \in C',t=M(s)}
\end{equation}

\begin{equation}
D'={(s',t')|s' \in C,t'=M'(s')}
\end{equation}

Based on these mappings, with $N=|D|+|D'|$, we can approximate the following metrics, first Hausdorff distance:

\begin{equation}
H_D(C,C')=\max\left\{\max_{(s,t) \in D}\|s-t\|,\max_{(s',t') \in D'}\|s'-t'\|\right\}
\end{equation}

and then root mean squared distance:

\begin{equation}
RMS_D(C,C')=\sqrt{\frac{1}{N}\left(\sum_{(s,t) \in D}\|s-t\|^2+\sum_{(s',t') \in D'}\|s'-t'\|^2\right)}
\end{equation}

Note that we do not evaluate distance bilaterally as our sampling algorithm requires manifold closed curves, whereas the reconstructed curves in our experiments may be open and non-manifold, and therefore measuring distance from a sampling on them would be less meaningful.

\subsection{Benchmark Driver and Test Scripts}\label{sec:b_script}
\emph{CurveBenchmark.cpp} is the C++ driver program of the curve reconstruction benchmark. This program consists of an algorithm list and functions for input-output processing, sampling, noise/outlier synthesis and evaluation. The driver program can be run from the terminal or using a test script. While running the driver executable, necessary arguments along with the command line options should be provided to interpret the argument type or values. A few examples of arguments and associated options are input file ($-i$), output file ($-o$), algorithm name ($-a$), and ground truth file ($-g$). The parameter $-h$ displays the list of options and their usage.

The architecture of our benchmarks consists of some test scripts that quantitatively and qualitatively evaluate the curve reconstruction algorithms by feeding the input data sets to algorithms and processing the evaluated data into graphs. Each test script includes a list of algorithms that need to be considered for the experiment and a list of test data. To add a new reconstruction algorithm, namely $A$, for a particular curve feature, $A$ has to be first downloaded to the benchmark repository, compiled and the executable of $A$ has to be linked to the benchmark driver via the given make file. Additionally, $A$ needs to be included in the algorithm list of the benchmark driver program. Finally, $A$ has to be included in the appropriate test script and to be run, which in turn invokes the benchmark driver, thereby generating the resulting polygonal curves as well as the graph plots. All this can simply be copied and modified from the existing structure. It should be noted that our test scripts also make it easy for the benchmark users to run selective experiments using a subset of the benchmark data. The list of test scripts in our benchmark evaluates RMS error for the following input data unless otherwise noted:
\begin{itemize}
  \item \textbf{run-sampling.sh:} $\epsilon$-sampled~\cite{amenta98curve} test data
  \item \textbf{run-noisy.sh:} perturbed with uniform noise
  \item \textbf{run-lfsnoise.sh:} perturbed with lfs-based noise
  \item \textbf{run-outliers.sh:} added outlier points
  \item \textbf{run-manifold.sh:} whether reconstruction is a manifold
  \item \textbf{run-sharp-corners.sh:} sharp feature curves
  \item  \textbf{run-open-curves.sh:} open curves
  \item \textbf{run-multiple-curves.sh:} multiply connected curves
  \item \textbf{run-intersecting.sh:} curves with intersections
\end{itemize}

\section{Evaluation \& Results}
In this section, we demonstrate the utility of the benchmark by comparing the curve reconstruction algorithms included in the benchmark. The algorithms have been evaluated on different test data using various criteria implemented in the benchmark. For the sake of fair comparison, a majority of the experiments use only the interpolatory algorithms, i.e., optimal transport~\cite{degoes2011optimal} has been discussed separately. Both the quantitative and qualitative comparisons have been presented along with a detailed interpretation of the results. Besides the comparison of classic and recent algorithms, this section also focuses on the demonstration of a comprehensive experimental design for curve reconstruction techniques, i.e., to show how the quantitative and qualitative comparisons are designed and performed using the test data and the scripts available in the curve reconstruction benchmark. Note that any new curve reconstruction algorithm can be easily added to the benchmark by duplicating and adapting the appropriate wrappers and subsequently be compared with the existing algorithms in the benchmark.

\subsection{Quantitative Evaluation}
We rank the 14 algorithms by the following six aspects.
The closeness to the original is measured by computing the root of the mean-squared error (RMSE) metric.
This permits determining the robustness with respect to various sampling artifacts.

\begin{itemize}
	\item Sampling density as $\epsilon$-sampling
	\item Noise robustness as $\delta$ of bounding box diagonal
	\item Noise robustness as $\delta$ of lfs
	\item Noise+sampling density as $\epsilon$-sampling and $\delta$ of lfs
	\item Outliers robustness in \% of samples
	\item Average runtimes (in s)
\end{itemize}

The test set consists of point sets sampled from multiply connected, and disconnected curves and curves with sharp corners. For some experiments, we used smooth curves of the bunny sampled as required (see Figures \ref{fig:eval-sampling}, \ref{fig:eval-lfsnoise} and \ref{fig:eval-noiselfs}).

\subsubsection{Sampling Density}

\begin{figure}[h]
	\centering
	\includegraphics[width=3in]{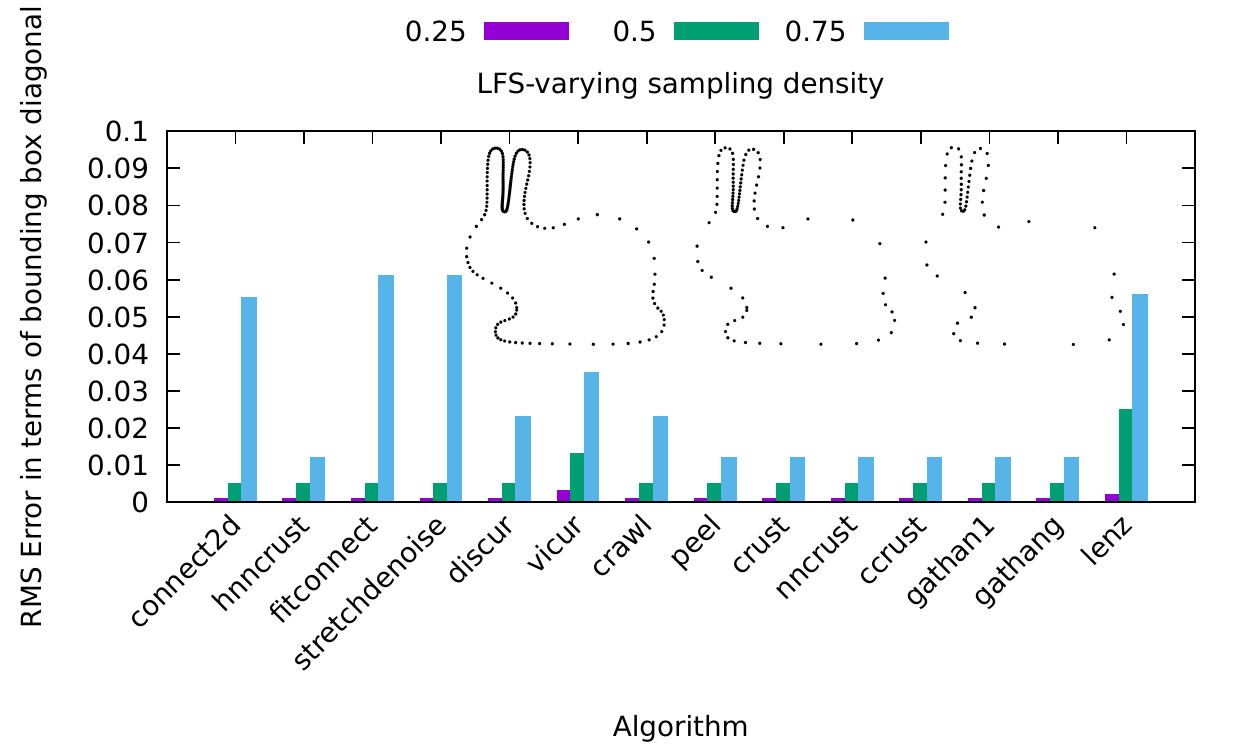}
	\caption{RMS Error of reconstructed curves from ground truth for a cubic B{\'e}zier curve sampled with $\epsilon=0.25, 0.5$ and $0.75$ ({\em run-sampling.sh}). The point sets sampled from the bunny curve are shown in the figure. }
	\label{fig:eval-sampling}
\end{figure}

First, we look at simple reconstruction from a non-uniform sampling of curves, free of artifacts such as noise or outliers.
For this, we determine an $\epsilon$-sampling on a cubic B{\'e}zier curve (bunny), see Subsection~\ref{sec:sampling} for our detailed implementation.
Reconstruction from sufficiently dense samples is not a difficult task.
In order to show which algorithms also work well on sparser sampling, we sample with decreasing density: $\epsilon=0.25, 0.5, 0.75$ (see Figure~\ref{fig:eval-sampling}).  As can be seen, irrespective of the density, for dense sampling ($\epsilon$ = 0.25,0.5), all the algorithms perform equally well (except for {\scshape Lenz} - which gives a comparatively poor performance in all cases, and {\scshape VICUR}). But, as the sampling becomes sparse ($\epsilon$ = 0.75), the performance of algorithms like {\scshape connect2D}, {\scshape fitconnect}, and {\scshape stretchdenoise} deteriorate compared to other algorithms. Note that minimum error levels stem from comparing the coarser reconstruction to the original smooth curve.
\subsubsection{ Robustness to Noise}

To evaluate the performance of the algorithms on noisy point set, we ran our benchmark on point sets perturbed by different noise levels. In our first experiment, we introduce uniform noise, then, we vary the noise in terms of the local feature size, and finally we fix the latter while varying the sampling density (also in terms of the lfs):

\textbf{Uniform Noise:}

\begin{figure}[h]
	\centering
	\includegraphics[width=3in]{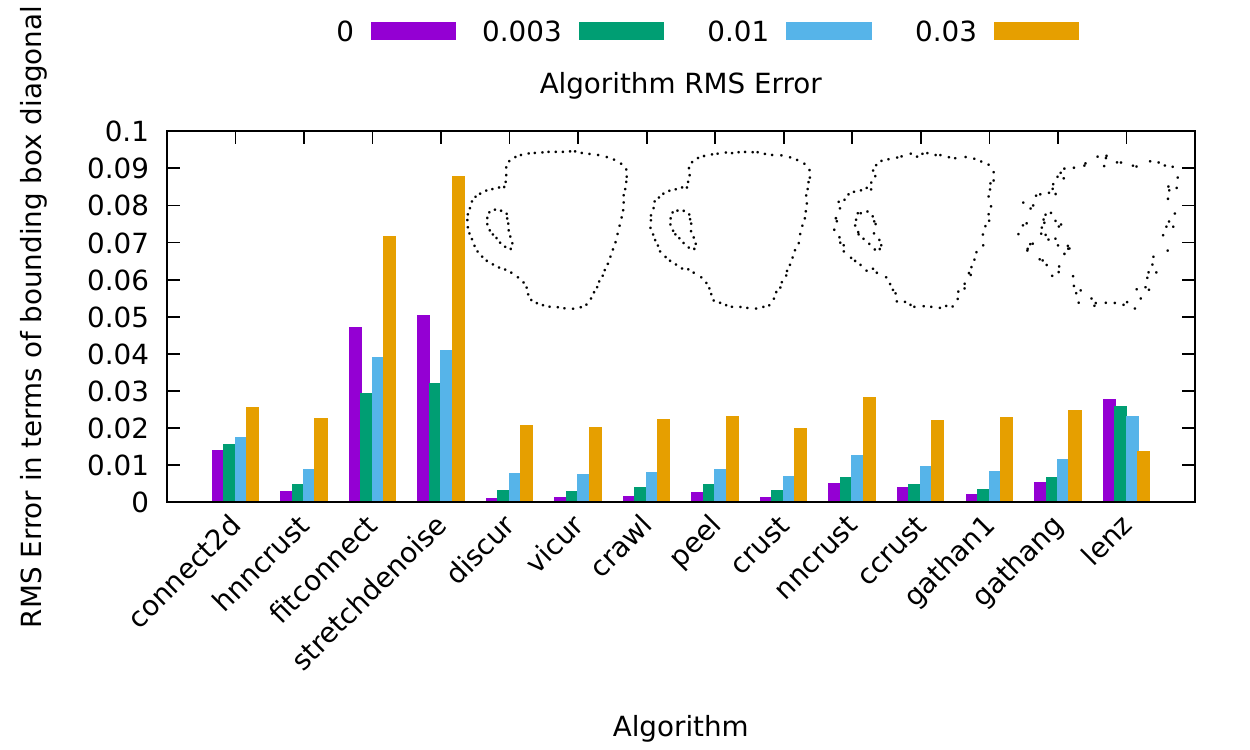}
	\caption{RMS Error of reconstructed curves from ground truth for the point sets used in Figure~\ref{fig:eval-exact}, with the points perturbed with {\em uniform} noise of $\delta=0.003, 0.01$ and $0.03$ as well as the non-noisy original ({\em run-noisy.sh}). An example input with varying noise is also shown in figure. {\scshape Lenz} results are not directly comparable as it reconstructs a non-manifold and thus largely the original as a subset.}
	\label{fig:eval-noisy}
\end{figure}

We simulate noise of maximum extent $\delta$ by defining perturbing a sample on the curve $s_i$ to $s_i'=s_i+\sigma v_i$, where $\sigma$ is an uniform random variable in $[0,\delta]$, $\delta$ is in terms of the bounding box diagonal, and $v_i$ is a unit vector of uniformly random direction, similar to the noise model used here~\cite{mehra2010visibility}.

Using different perturbation levels of 0.003, 0.01, 0.03, we generated 3 additional noisy datasets from the original for the 25 {\scshape Classic}-{\em manifold} points sets. Figure~\ref{fig:eval-noisy} shows a graph representing the performance of the various algorithms on our synthetically generated noisy dataset together with the non-noisy original. For the latter, reconstruction often fails already, the close plots show how it degrades with additional noise, except for {\scshape Lenz} which reconstructs less wrong cross-connections.
HVS- and Delaunay-based algorithms such as {\scshape crawl}, {\scshape peel}, {\scshape crust} families gave the best performance for uniformly distributed noise, while {\scshape Connect2D}, {\scshape FitConnect} and {\scshape StretchDenoise} fail to reconstruct the original in several cases due to sharp corners or too sparse sampling, but degrade much less for correctly reconstructed ones.

\textbf{LFS-based Noise:}

\begin{figure}[h]
	\centering
	\includegraphics[width=3in]{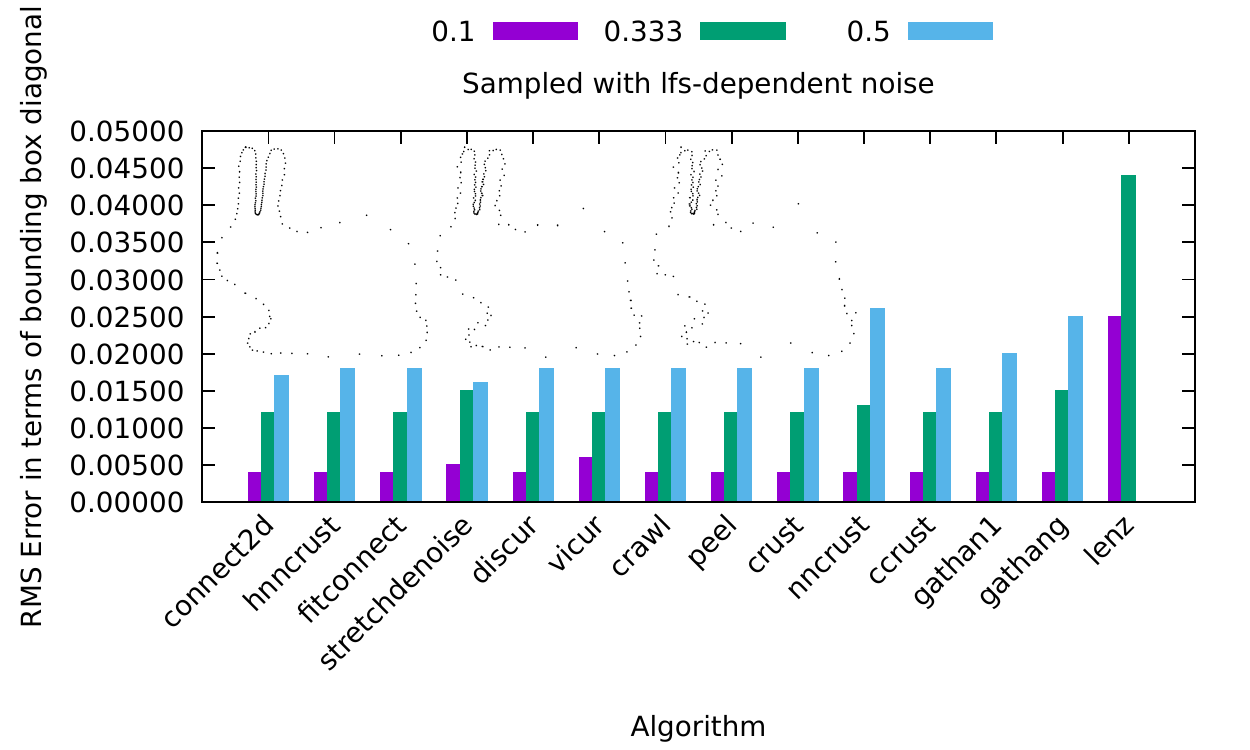}
	\caption{RMS Error of reconstructed curves from ground truth for the sets of cubic B{\'e}zier curves in Figure~\ref{fig:eval-sampling}, sampled with $\epsilon=0.3$, and the points perturbed with {\em local feature sized} noise of $\delta=0.1, \frac{1}{3}$ and $0.5$ ({\em run-lfsnoise.sh}). The figure also shows the lfs sampled bunny curve with varying noise ({\scshape Lenz} crashes for $\delta=0.5$, therefore the false 0 value).}
	\label{fig:eval-lfsnoise}
\end{figure}

In this experiment, we simulate noise in terms of the extent of the \emph{local feature size}. To add noise to such a sampling of the bunny as above, we perturb the sample $s_i$ only along the curve unit normal vector $n_i$ in order to preserve the sampling density as well as possible, such that $s_i'=s_i+\sigma n_i$, with uniform random $\sigma=[-\delta,+\delta]\mbox{lfs}(s)$. Figure~\ref{fig:eval-lfsnoise} shows the performance of the various curve reconstruction algorithms compared to the noise-free original, all with an $\epsilon$-sampling of $\epsilon=0.3$. The algorithms show similar characteristics as for global uniform noise as shown above. Algorithms such as {\scshape Connect2D}, {\scshape fitconnect} and {\scshape stretchdenoise} showed superior performance under the lfs-noise model, similar to the global uniform noise model tested above. This is no surprise as the algorithms such as {\scshape fitconnect} and {\scshape stretchdenoise} are especially designed to handle noise and provide guarantees under the \emph{lfs} based sampling.

\textbf{Varying Sampling Density + LFS Noise:}

\begin{figure}[h]
	\centering
	\includegraphics[width=3in]{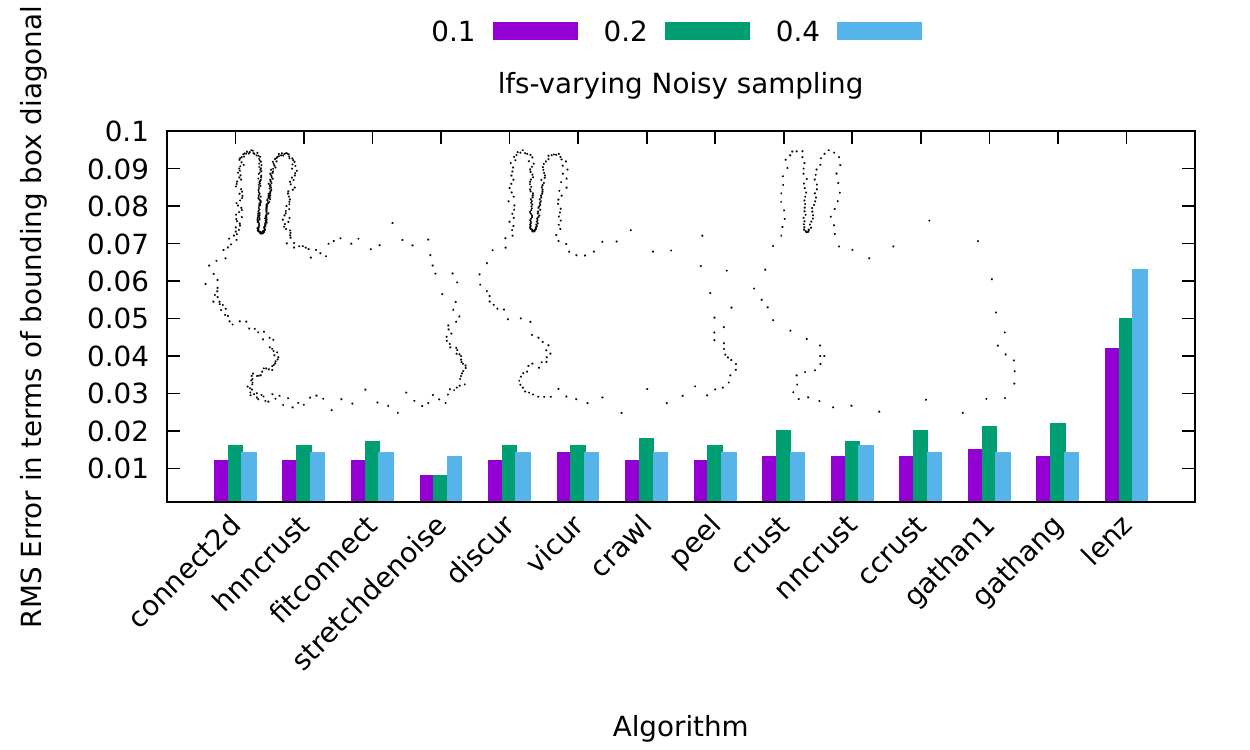}
	\caption{RMS Error of reconstructed curves from ground truth for sets of cubic B{\'e}zier curves, with the points perturbed with noise of $\delta=\frac{1}{3}$ and sampled with $\epsilon=0.1, 0.2$ and $0.4$ ({\em run-sampling-noise.sh}). The point sets of the bunny in the figure represent the inputs with varying sampling and noise.}
	\label{fig:eval-noiselfs}
\end{figure}

Now we introduce noise as above but with fixed $\delta=\frac{1}{3}$ and vary the sampling density with $\epsilon=0.1, 0.2, 0.4$ on the bunny, again according to Subsection~\ref{sec:sampling}.
We can see in Figure~\ref{fig:eval-noiselfs} that varying sampling density does not have a large impact on reconstruction from noisy samples across all algorithms.

\subsubsection{Outliers}

\begin{figure}[h]
	\centering
	\includegraphics[width=3in]{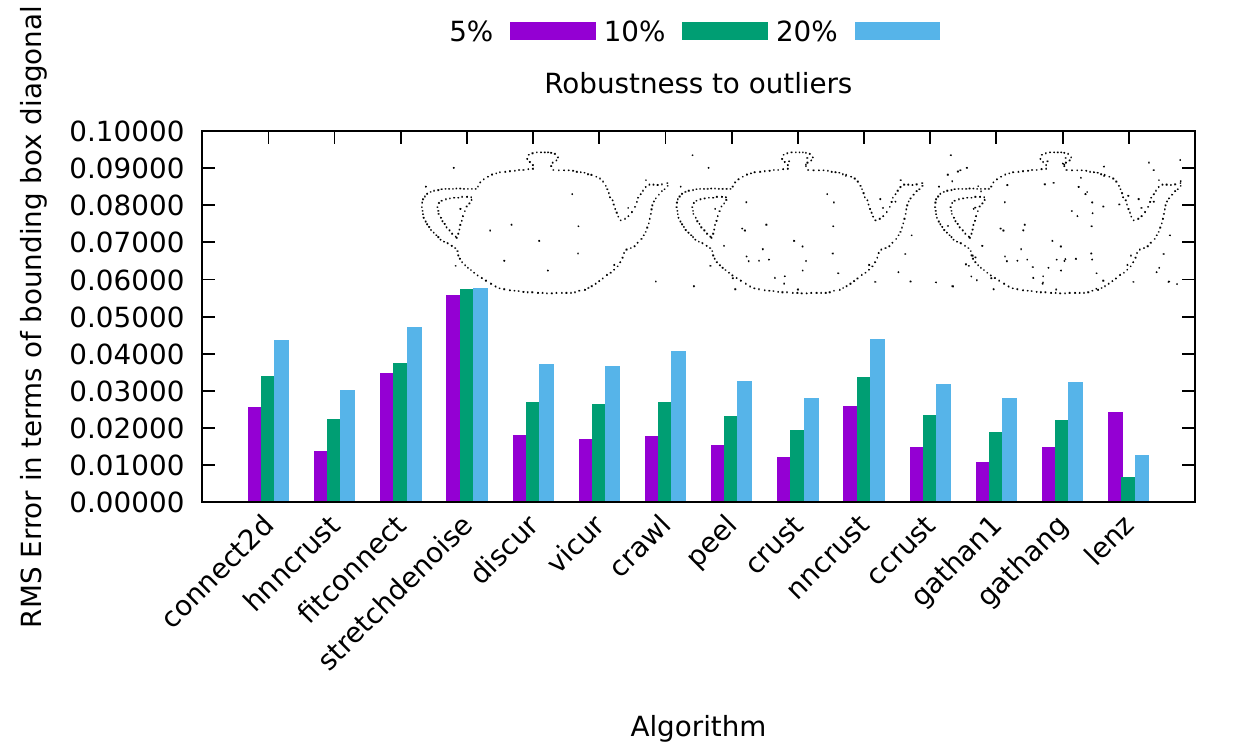}
	\caption{RMS Error of closed reconstructed curves for the classic point sets used in Figure~\ref{fig:eval-exact}, with a share of $5\%, 10\%$ and $20\%$ outliers added. ({\em run-outliers.sh}). {\scshape Lenz} results are not directly comparable as it reconstructs a non-manifold and thus largely the original, while connecting outliers.}
	\label{fig:eval-outliers}
\end{figure}

To evaluate the performance of reconstruction in the presence of outliers, we simply generate $n\%$ additional samples uniformly distributed inside the bounding box of the point set. All the interpolatory algorithms were tested on 25 input curves from {\scshape Classic} dataset. Figure~\ref{fig:eval-outliers} shows the robustness of reconstruction after adding a varying amount of outliers as uniformly random distributed points inside the bounding box of the input point set. While {\scshape crust} and {\scshape HNNCrust} algorithms perform very well in the presence of outliers, other algorithms such as {\scshape StretchDeNoise} and {\scshape FitConnect} are found to be comparatively sensitive to outliers.

\begin{figure}[h]
	\centering
	\includegraphics[width=3in]{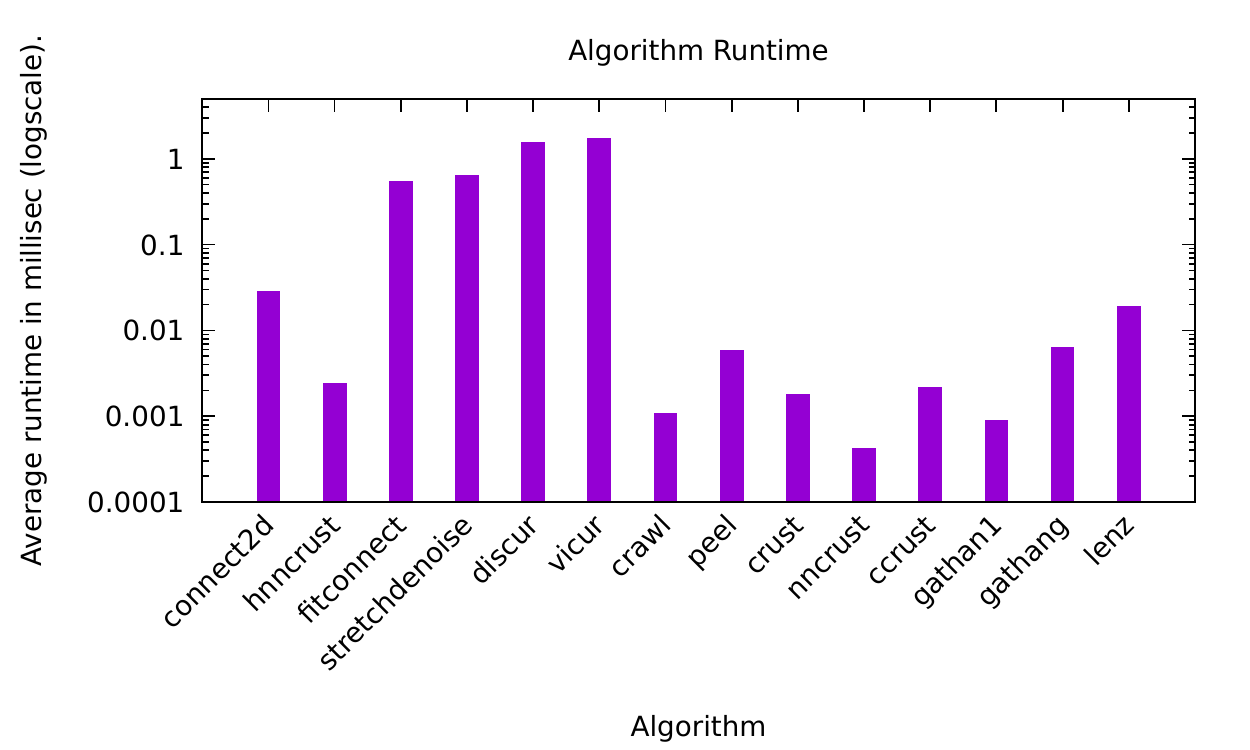}
	\caption{Averaged runtime for the reconstruction of 2183 manifold point sets for 14 algorithms ({\em run-manifold.sh}). Note the log scale due to the large variance.}
	\label{fig:eval-runtime}
\end{figure}
\begin{table}
	\begin{center}
		\begin{tabular}{|l|r|}
			\hline
			Algorithm & Runtime(s) \\
			\hline
			{\scshape Crust}~\cite{amenta98curve} & 0.0018 \\
			{\scshape NN-Crust}~\cite{dey99curve} & 0.0004 \\
			{\scshape CCrust}~\cite{dey99conservative} & 0.0022 \\
			{\scshape Gathan1}~\cite{dey01corners} & 0.0009 \\
			{\scshape GathanG}~\cite{dey02gathang} & 0.0066 \\
			{\scshape Connect2D}~\cite{ohrhallinger13connect2d} &0.0277 \\
			{\scshape Crawl}~\cite{parakkat2016crawl} & 0.0012 \\	
	    		{\scshape HNN-Crust}~\cite{ohrhallinger2016hnn} & 0.0023 \\
			{\scshape FitConnect}~\cite{ohrhallinger2018fitconnect} & 0.5081 \\
			{\scshape StretchDenoise}~\cite{ohrhallinger2018stretchdenoise} & 0.6182 \\
			{\scshape Peel}~\cite{parakkat2018peeling} & 0.0058 \\			
			{\scshape DISCUR}~\cite{zeng08distance} & 1.5291 \\
			{\scshape VICUR}~\cite{nguyen08vicur} & 1.7037 \\
            		{\scshape Lenz}~\cite{lenz06curve} & 0.0576\\
			\hline
		\end{tabular}
		\caption{Average runtime of 14 algorithms reported for 2183 manifold point sets.}
		\label{table:run-time}
	\end{center}
\end{table}

\subsubsection{ Computational Time}\label{sec:runtime}
Figure~\ref{fig:eval-runtime} shows a bar chart of average run times for the various curve reconstruction algorithms. 
The test set consists of 2183 noise-free manifold point sets with an average of 283.9 points per test case, each of which represents a closed boundary of either a silhouette image or a geometric shape. We use {\scshape Classic} as well as {\scshape Image} contours for the experiment. Nearly 99\% of the inputs were extracted from silhouette images available in different image databases mentioned in Section \ref{sec:data}. The exact time values are also reported in Table~\ref{table:run-time}. Computational times are based on the experiment performed on an Intel core-i7 2.4 GHz CPU with 16 GB memory. As indicated by the bar chart and the Table, {\scshape nncrust} is the fastest of all the compared algorithms with an average time of 0.0004 ms for reconstructing 2183 noise-free point sets, contrasting to 1.78s for slowest {\scshape Vicur}, with the majority of algorithms finishing in <10ms.

\subsection{Qualitative Comparison}

We compare how well the algorithms perform on these qualitative aspects:
\begin{itemize}
	\item Manifold curves
\item Non-manifold
	\item Sharp corners
	\item Open curves
	\item Multiple curves
\end{itemize}

\begin{figure*}[ht]
	\centering
	\includegraphics[width=16cm]{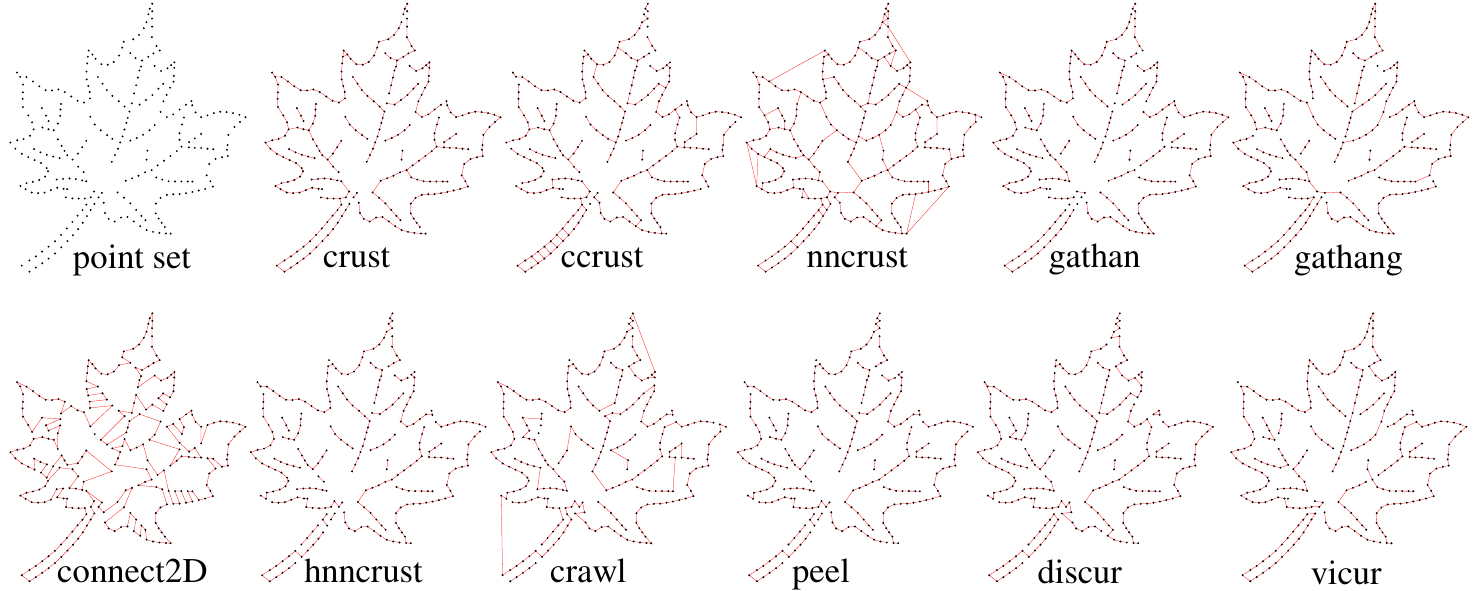}
	\caption{A qualitative comparison of different algorithms on the leaf input that contains sharp-corners, non-manifold edges, multiple and open curves ({\em run-qualitative.sh}).}
	\label{fig:eval-nsom}
\end{figure*}

Figure \ref{fig:eval-nsom} shows example reconstruction results by the 11 algorithms which are able to handle all these aspects contained in the leaf input (sharp-corners, non-manifold edges, multiple and open curves).

\subsubsection{Manifold Curves}

 \begin{figure}[h]
	\centering
	\includegraphics[width=3in]{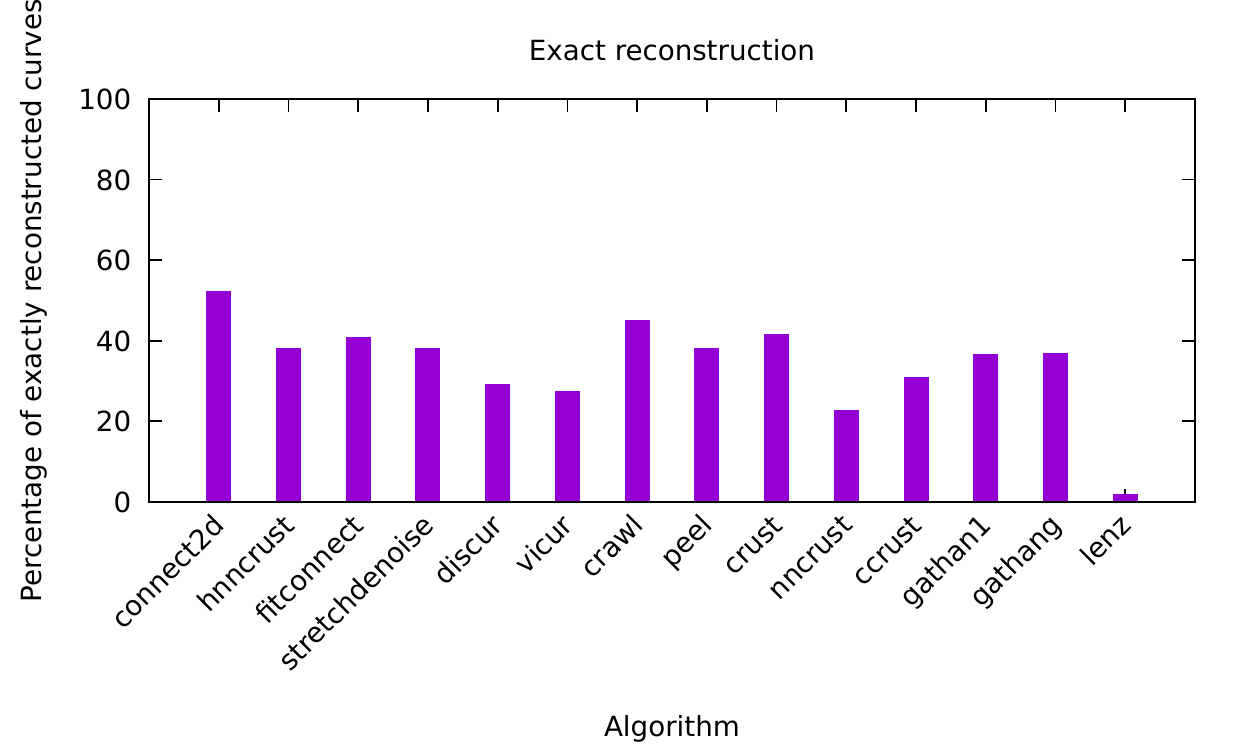}
	\caption{Percentage of exactly reconstructed curves from a set of 2183 noise-free point sets predominantly generated from images, for 14 algorithms ({\em run-manifold.sh}).}
	\label{fig:eval-exact}
\end{figure}

We evaluate the reconstruction algorithms for their performance on the manifold curves. The set of 14 algorithms considered for this experiment were run on 2183 noise-free point sets used for the run time experiment (Section \ref{sec:runtime}). Figure~\ref{fig:eval-exact} shows a qualitative performance of the algorithms in exact reconstruction of manifold curves. The percentage of exactly reconstructed manifold curves lies in the range of 22.6-45\% for the majority of algorithms. Among the tested algorithms, {\scshape Connect2D} performs best, with 52.13\% of curves reconstructed faithfully. Contrary to the design objectives of {\scshape lenz}, only 1.7\% of the manifold curves were reconstructed correctly. This could be partly due to the presence of holes or multiply connected curves as we observed that {\scshape lenz} algorithm does not handle multiple curves well.

\subsubsection{ Non-manifold Point Sets}

\begin{figure}[h]
	\centering
	\includegraphics[width=3in]{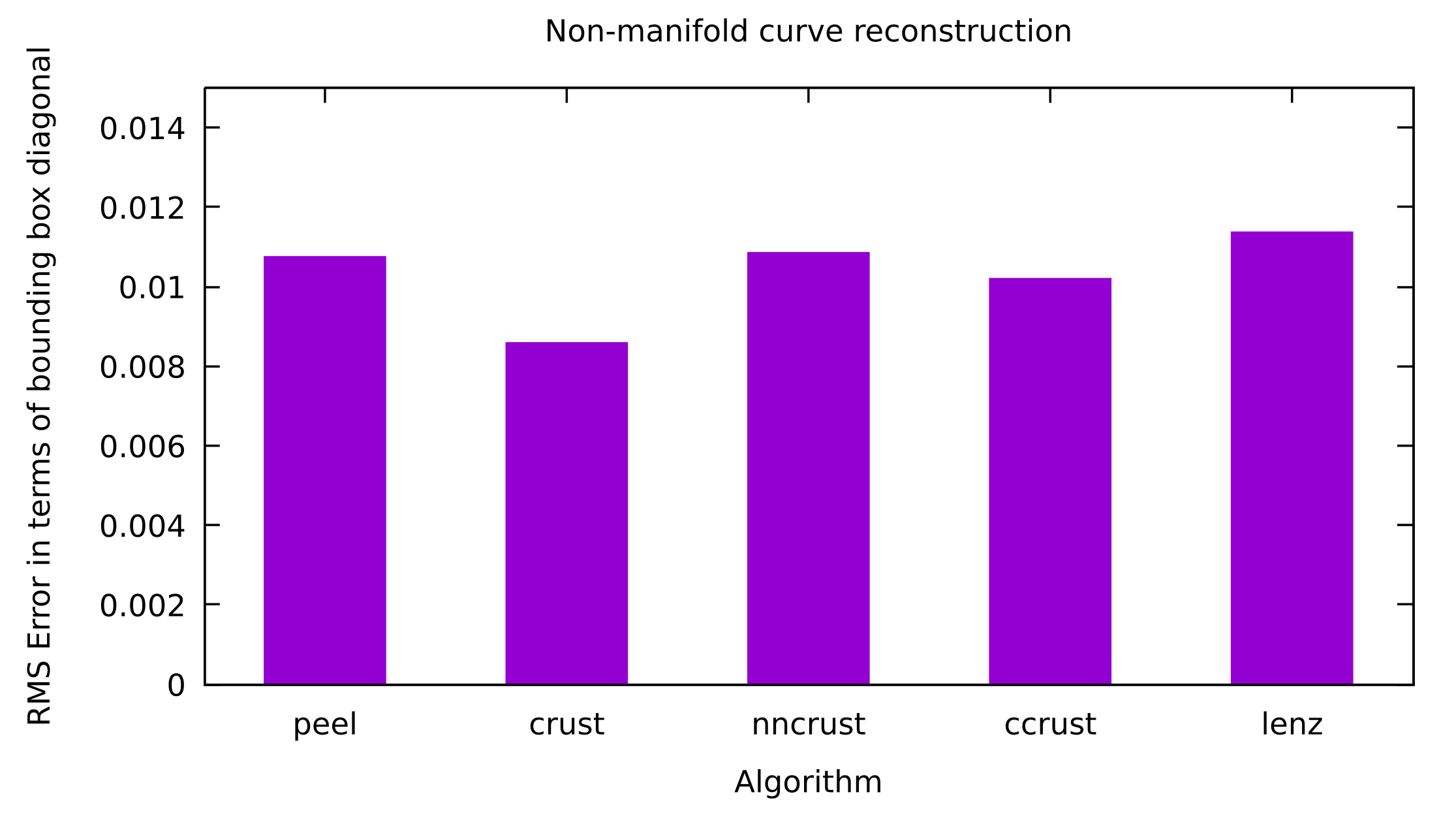}
	\caption{RMS error of reconstructed non-manifold curves compared with ground truth. We tested these algorithms on 26 inputs which consists of 16 curves collected from various curve reconstruction papers and 10 synthetically created curves ({\em run-non-manifold.sh}).}
	\label{fig:eval-non-manifold}
\end{figure}

Figure \ref{fig:eval-non-manifold} shows the bar chart of our experimentation on non-manifold curves. In this experiment, we have included only five algorithms that are known to handle non-manifold curves. We used 16 point sets created from various papers and ten synthetic point sets for the experiments. The synthetic point sets were generated by sampling randomly drawn self-intersecting curves. As can be seen, all the five algorithms gave a similar performance on non-manifold curves in terms of RMS error. In terms of the percentage of correct reconstruction, approximately 11\% of the curves were faithfully reconstructed, which is the maximum by any of the tested algorithms.  From the results, we observed that all the algorithms except {\scshape Lenz} reconstructed the curves reasonably well. However, the resulting curves lacked one/or two edges and therefore, did not exactly match the corresponding groundtruths. This could potentially lead to such a low percentage of correct reconstruction.

\begin{figure}[h]
	\centering
	\includegraphics[width=3in]{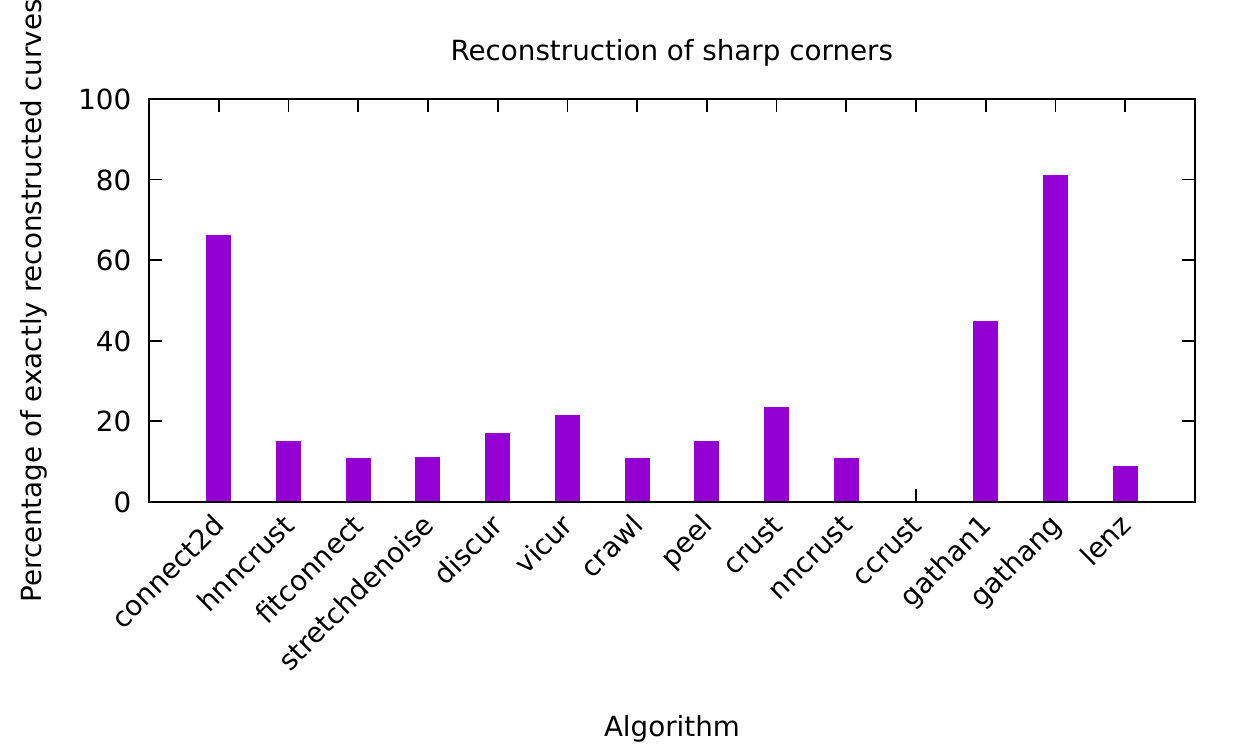}
	\caption{Percentage of exactly reconstructed curves having sharp corners by various algorithms. We tested the algorithms on 47 inputs which consists of 21 curves collected from various curve reconstruction papers and 26 synthetically created curves ({\em run-sharp-corners.sh}).}
	\label{fig:eval-sharp-corners}
\end{figure}
\begin{figure}[h]
	\centering
	\includegraphics[width=3in]{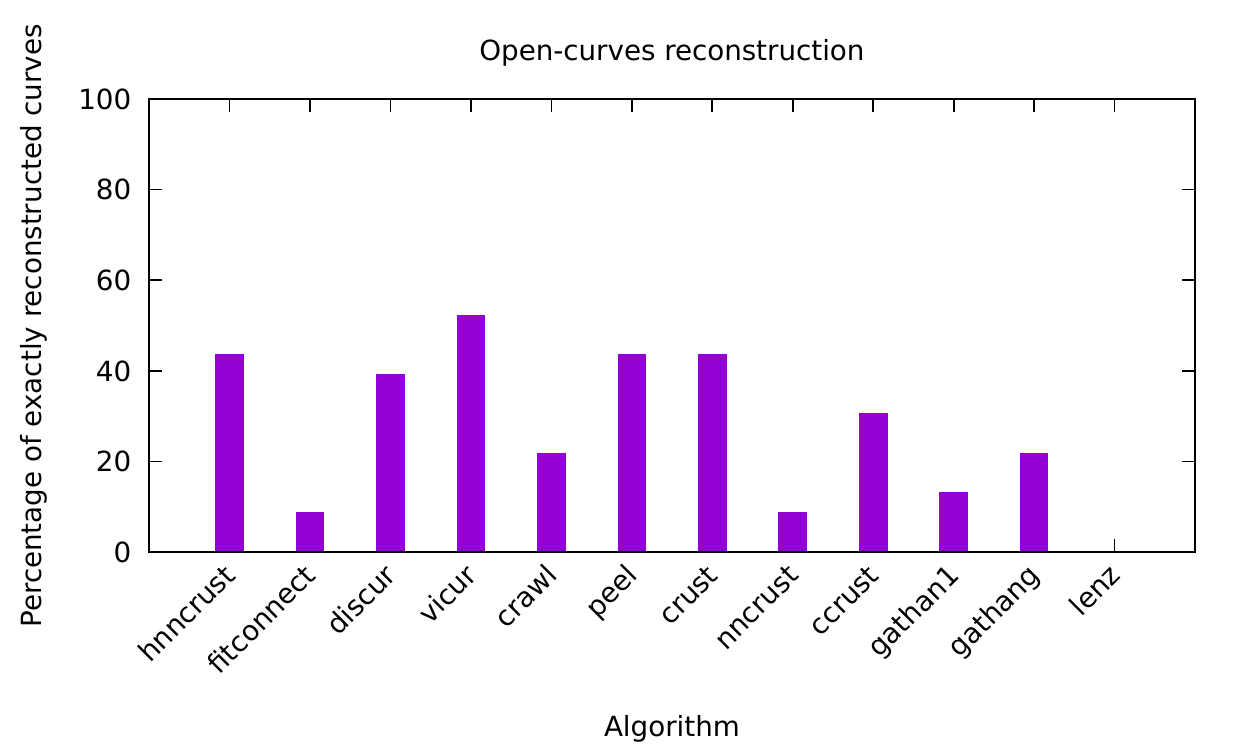}
	\caption{Percentage of exact reconstruction of open curves by different algorithms. We tested the algorithms on 23 different open curves collected from various curve reconstruction papers ({\em run-open-curves.sh}).}
	\label{fig:eval-open-curves}
\end{figure}
\begin{figure}[h]
	\centering
	\includegraphics[width=3in]{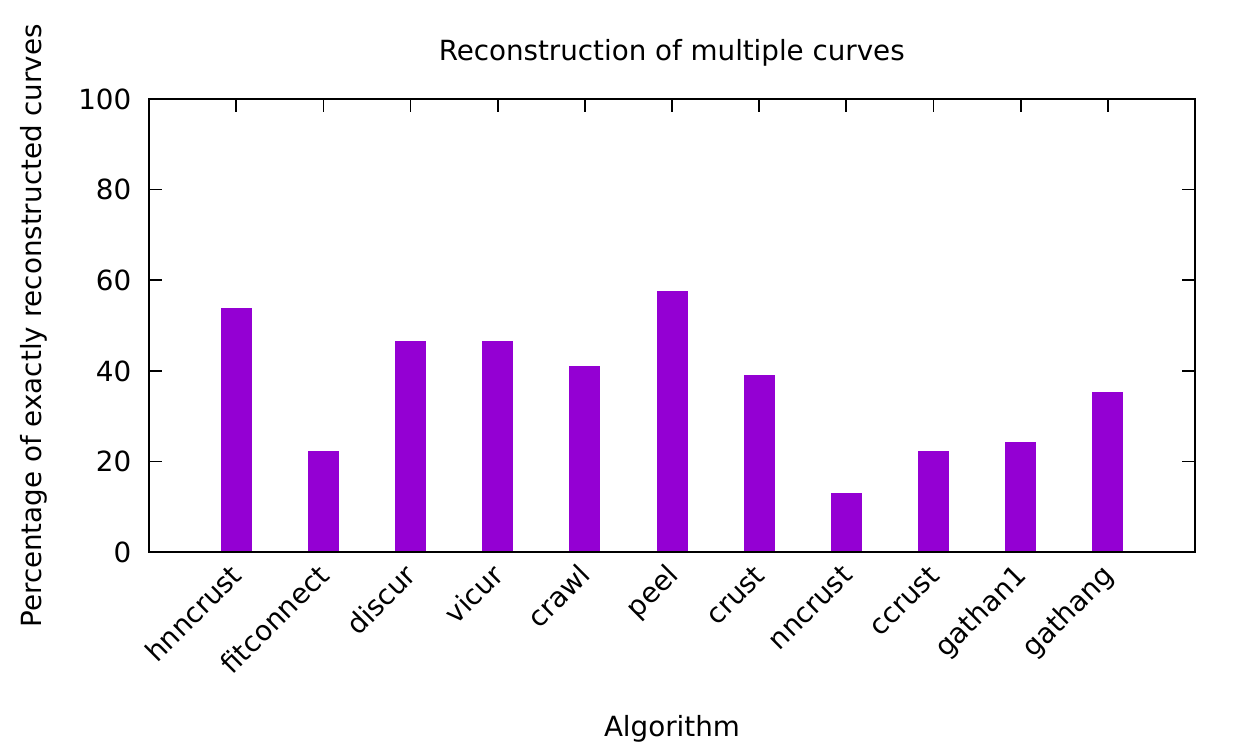}
	\caption{Percentage of exactly reconstructed curves (multiply connected as well as disconnected curves) by various algorithms. We tested the algorithms on 52 different point sets collected from various curve reconstruction papers and 2 point sets sampled from image silhouette boundaries ({\em run-multiple-curves.sh}).}
	\label{fig:eval-multi-curves}
\end{figure}

\subsubsection{Sharp Corners} Figure~\ref{fig:eval-sharp-corners} shows the bar chart representing our experiments on curves with sharp corners. We tested algorithms which handle sharp corners on a test set consisting of 47 input curves. The test set consists of classic curves collected from the literature and synthetic data. Each curve in the synthetic dataset is a closed convex curve which was generated using an arc of a unit circle opening to two tangent lines joining at a specified angle to form a sharp corner. Circles were sampled with 10, 16 and 20 poins with opening angles of the tangents of 10-90 degrees in steps of 10 and 5, and sampled with similar density. {\scshape GathanG} performs best, it was specifically designed for curves with sharp features, and correctly reconstructs 80.85\% of the test cases. The second best algorithm in handling sharp corners is {\scshape Connect2D} which handled 65.95\% of the inputs correctly, followed by {\scshape Gathan1} with a success rate of 44.68\%. The remaining algorithms succeed only for few cases, i.e., [0, 23.40]\%.

\subsubsection{Open Curves} Figure~\ref{fig:eval-open-curves} shows a bar diagram with the percentage of correctly reconstructed open curves per algorithm along the y-axis. Input to the algorithms were a set of 23 open curve point sets manually selected from various reconstruction papers. We excluded {\scshape connect2D} and {\scshape stretchdenoise} because both these algorithms were not designed for open curves. The {\scshape vicur} algorithm reconstruct most (52.2\%) of the open curves . Other algorithms such as {\scshape hnncrust}, {\scshape crust}, {\scshape peel} and {\scshape discur} also perform reasonably well on open curves.
\begin{figure*}[!htbp]
	\centering
	\includegraphics[width=16cm]{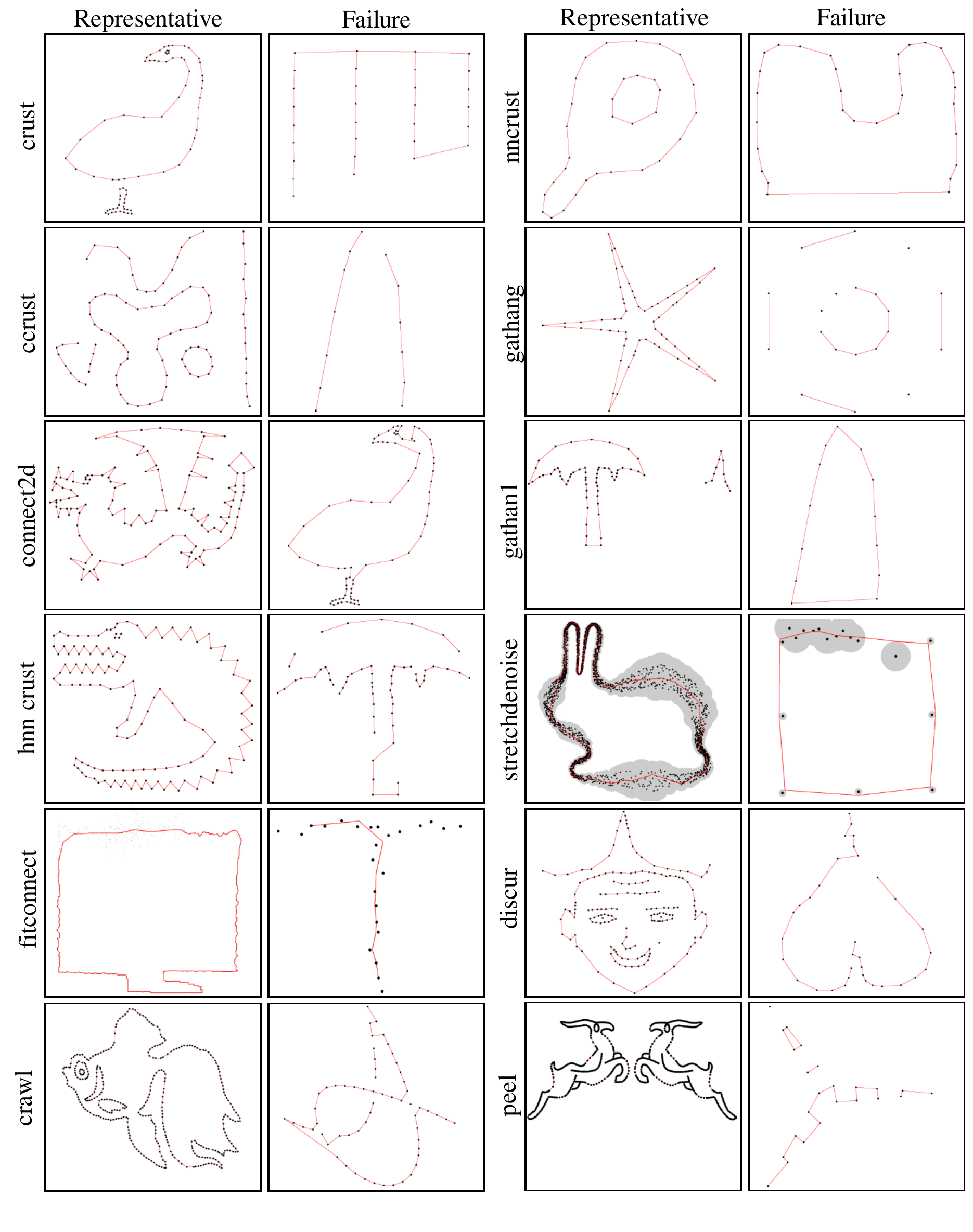}
	\caption{Representative and failed results of various curve reconstruction algorithms (part 1).}
	\label{fig:eval-rep}
\end{figure*}

\subsubsection{Multiple Curves} We considered two types of inputs in this experiment: multiply connected curves (curves with holes) and disconnected curves. We evaluated 11 algorithms on 54 point sets. We omitted {\scshape connect2D}, {\scshape stretchdenoise} and {\scshape lenz} because these algorithms are not designed to handle multiple curves. Figure~\ref{fig:eval-multi-curves} shows the bar diagram displaying the percentage of curves reconstructed per algorithm along the y-axis. In our experiment, {\scshape peel} and {\scshape hnn-crust} handled multiple curves comparatively well.

\begin{figure}[!h]
	\centering
	\includegraphics[width=8cm]{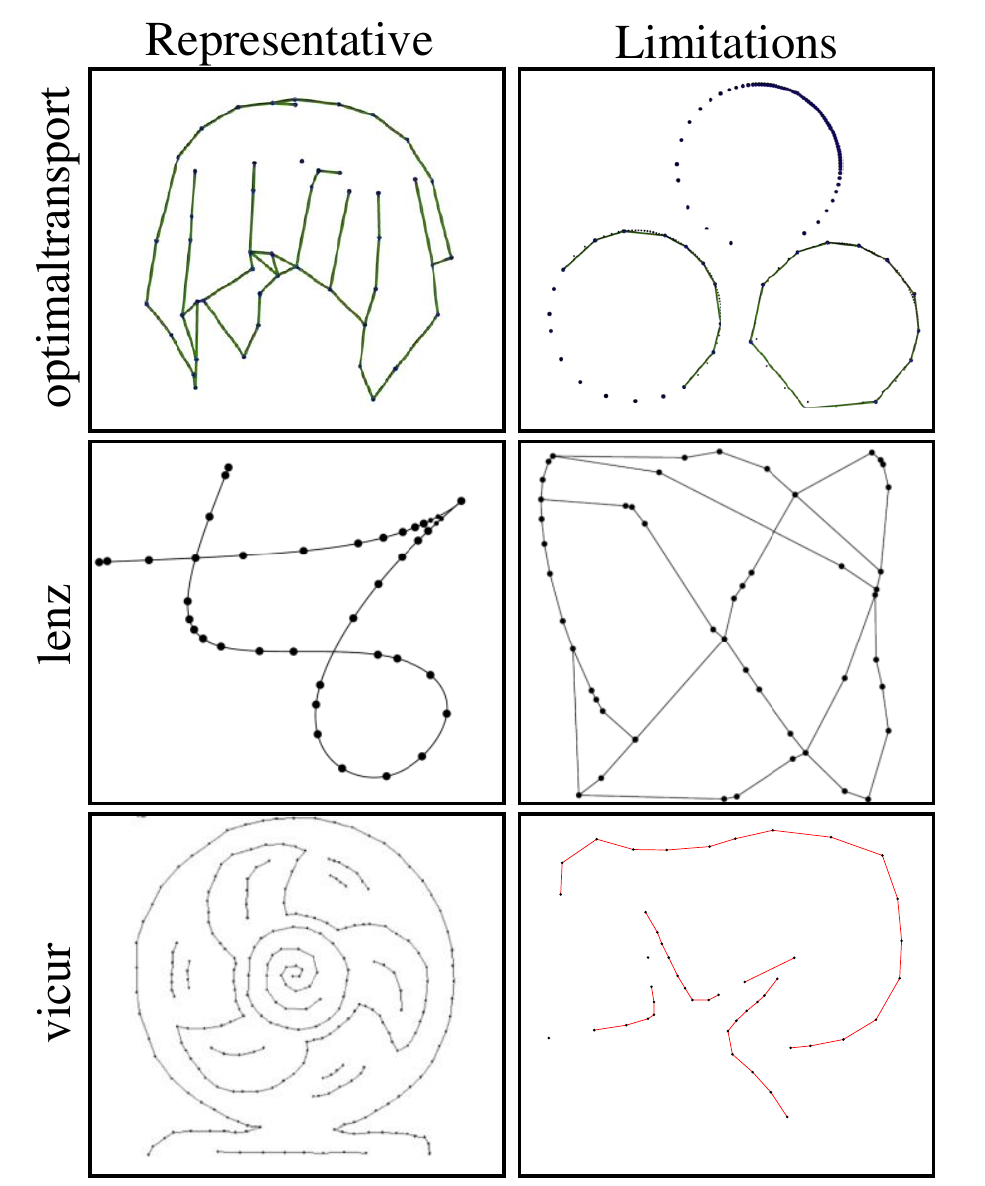}
	\caption{Representative and failed results (part 2) of {\scshape OptimalTransport} (row 1), {\scshape Lenz} (row 2) and {\scshape VICUR} (row 3). Image courtesy: row 1~\cite{degoes2011optimal}, row 2~\cite{lenz06curve} and row 3 (left)~\cite{nguyen08vicur}. Note that all three algorithms are sensitive to lower sampling density. }
	\label{fig:eval-rep2}
\end{figure}

\subsection{Guarantees as Sampling Conditions}

\begin{table}[!h]
	\begin{center}
		\begin{tabular}{|l|r|}
			\hline
			Algorithm & Sampling condition \\
			\hline
			{\scshape Crust}~\cite{amenta98curve} & $\epsilon<0.252$ \\
			{\scshape NN-Crust}~\cite{dey99curve} & $\epsilon<\frac{1}{3}$ \\
			{\scshape CCrust}~\cite{dey99conservative} & $\epsilon<0.0625$ \\
			{\scshape Gathan}~\cite{dey01corners} & none \\
			{\scshape GathanG}~\cite{dey02gathang} & $\epsilon<0.5$, $\alpha>150^\circ$ \\
			{\scshape Lenz}~\cite{lenz06curve} & $\epsilon<0.48$ (no proof) \\
			{\scshape Connect2D}~\cite{ohrhallinger13connect2d} & $\epsilon<\frac{1}{2}$, $u<1.609$ \\
			{\scshape Crawl}~\cite{parakkat2016crawl} & none \\
            {\scshape HNN-Crust}~\cite{ohrhallinger2016hnn} & $\epsilon<0.47 \equiv \rho<0.9$ \\ 
			{\scshape Fitconnect}~\cite{ohrhallinger2018fitconnect} & as {\scshape HNN-Crust} \\
			{\scshape Stretchdenoise}~\cite{ohrhallinger2018stretchdenoise} & as {\scshape HNN-Crust} \\
			{\scshape Peel}~\cite{parakkat2018peeling} & none \\
			{\scshape Opt.Trans.}~\cite{degoes2011optimal} & none \\
			{\scshape DISCUR}~\cite{zeng08distance} & vision function \\
			{\scshape VICUR}~\cite{nguyen08vicur} & none \\
			\hline
		\end{tabular}
		\caption{Sampling conditions of algorithms w.r.t. reconstruction of a manifold curve, where applicable. $\epsilon$ refers to $\epsilon$-sampling~\cite{amenta98curve}, $\alpha$ to an opening angle between adjacent edges, $u$ to local uniformity as factor of adjacent edge lengths.}
		\label{table:compared-conditions}
	\end{center}
\end{table}

For some algorithms, reconstruction is guaranteed if a certain sampling condition is fulfilled, this specifies e.g. with which maximum distance the points are allowed to be sampled on the original curve, often depending on some other criteria, but other conditions also exist.
Older algorithms like $\alpha$-shapes based Ball-pivoting~\cite{bernardini1997sampling} require a globally uniform maximum distance for samples.
Most of the algorithms compared here in Table~\ref{table:compared-conditions} rely on the $\epsilon$-sampling condition, which depends on the local size of features.
Among these, {\scshape HNN-Crust} performs best with $\epsilon<0.47$.
$\epsilon$-sampling can be relaxed if other conditions are added, such as an angle condition ({\scshape GathanG}) or a maximum factor between adjacent edge lengths ({\scshape Connect2D}).
Ohrhallinger et al.~\cite{ohrhallinger2016hnn} propose a new kind of sampling condition, $\rho$-sampling, which applies to curve intervals instead of curve points, and therefore permits a sparser sampling and thus less required points than the $\epsilon$-sampling they show it correlates with.
Funke et al. ~\cite{funke01curve} design a sampling condition explicitly for sharp corners (algorithm not compared here), but it is very complex and requires 8 parameters.
{\scshape DISCUR} uses a complex vision function as sampling condition.
We observed that often curves can be successfully reconstructed from much sparser sampling than proven for the algorithms, across the field. This indicates that sampling conditions could be still improved.

\subsection{Summary}
Figures \ref{fig:eval-rep} and \ref{fig:eval-rep2} show manually chosen examples of successful and failed reconstructions for all the 15 algorithms reported in the respective papers. Failure cases represent worst case point sets for various aspects such as noise-free, sparsely-sampled, noisy, outliers, noisy+outliers, non-manifold, sharp corners etc. We report the best two algorithms for handling specific types of input and curve characteristics in Table \ref{table:summary}.


\begin{table}[!h]
\small
	\begin{center}
		\begin{tabular}{|l|l|}
			\hline
			Curve/Input feature & Best two algorithms in order \\
			\hline
			Uniform Noise & {\scshape DISCUR}, {\scshape VICUR} \\
			Non-uniform Noise & {\scshape Stretchdenoise}, {\scshape Connect2D} \\
			Outliers & {\scshape HNN-Crust}, {\scshape Crust}\\	
			Non-uniform sampling & {\scshape HNN-Crust}, {\scshape Peel} \\
            		Runtime & {\scshape NN-Crust}, {\scshape Gathan1}\\
			Manifold curves & {\scshape Connect2D}, {\scshape Crawl}\\
            		Non-manifold curves & {\scshape Crust}, {\scshape Lenz} \\
			Sharp features & {\scshape GathanG}, {\scshape Connect2D} \\
			Open curves & {\scshape VICUR}, {\scshape HNN-Crust}\\	
    			Multiple curves & {\scshape Peel}, {\scshape HNN-Crust}  \\			
			\hline
		\end{tabular}
		\caption{Summary of the experimental study. We list the best and second-best algorithm in each category of input/curve characteristic. Note that a few algorithms were not included in these experiments due to their unavailability as open source and hence, the proposals that we make here are not comprehensive. The two fastest algorithms were reported based on the manifold curve reconstruction experiment on a test set consisting of 2183 inputs. }
		\label{table:summary}
	\end{center}
\end{table}

\section{Conclusion}

After analyzing the properties of 36 algorithms and comparing up to 15 of them, we showed that there are quite diverse approaches for curve reconstruction. This stems from the various challenges in the field, ranging from connecting uniformly distributed feature points over noisy silhouettes from sensors, non-smooth curves to non-uniform sampling and sketches which do not represent an outline.
In order to choose a suitable algorithm for a specific application, the user can first narrow the field by availability (as open source) and category, which roughly corresponds to the above applications, then select among those based on importance regarding the criteria compared in our evaluation. In our survey, we also describe the evolution of the algorithms, so some are simply outperformed by later ones.
In terms of theoretical guarantees, the potential of $\epsilon$-sampling only being based on local feature size seems to be maxed out.

\subsection{Future Directions}
In this mature field we still consider some directions to be worthwhile for future work, mostly building onto this fundamental research surveyed here.

\begin{itemize}
  \item Improving and simplifying sampling conditions, especially for non-smooth and self-intersecting curves
  \item Reconstructing curves from hand drawn sketches with varying stroke thickness and intensity
  \item Deep learning on curves, as for surface reconstruction
  \item Reconstruct smooth curves instead of (open) polygons
  \item Reconstruction of surfaces from networks of 3D curves
\end{itemize}


\bibliographystyle{eg-alpha}
\bibliography{star}

\newcommand{\etalchar}[1]{$^{#1}$}
\begin{thebibliography}{\uppercase{DGCSAD11}}

\bibitem[107]{1070}
The 1070-shape database.
\newblock \url{http://vision.lems.brown.edu/sites/default/files/1070db.tar.gz}.
\newblock [Online; accessed 19-October-2019].

\bibitem[AAA{\etalchar{*}}09]{aichholzer09medialaxis}
\textsc{Aichholzer O., Aigner W., Aurenhammer F., Hackl T., J{\"u}ttler B.,
  Rabl M.}:
\newblock Medial axis computation for planar free-form shapes.
\newblock \emph{Computer-Aided Design 41}, 5 (2009), 339 -- 349.
\newblock Voronoi Diagrams and their Applications.

\bibitem[ABCC]{url:concorde}
\textsc{Applegate D., Bixby R., Chvatal V., Cook W.}:
\newblock Concorde tsp solver.
\newblock \url{http://www.math.uwaterloo.ca/tsp/concorde.html}.
\newblock [Online; accessed 12-September-2018].

\bibitem[ABCo{\etalchar{*}}03]{Alexa03computingand}
\textsc{Alexa M., Behr J., Cohen-or D., Fleishman S., Levin D., Silva C.~T.}:
\newblock Computing and rendering point set surfaces.
\newblock \emph{IEEE Transactions on Visualization and Computer Graphics 9}
  (2003), 3--15.

\bibitem[ABE98]{amenta98curve}
\textsc{Amenta N., Bern M.~W., Eppstein D.}:
\newblock The crust and the beta-skeleton: Combinatorial curve reconstruction.
\newblock \emph{Graphical Models and Image Processing 60}, 2 (1998), 125--135.

\bibitem[ABE09]{attali09medialaxis}
\textsc{Attali D., Boissonnat J.-D., Edelsbrunner H.}:
\newblock Stability and computation of medial axes-a state-of-the-art report.
\newblock In \emph{Mathematical foundations of scientific visualization,
  computer graphics, and massive data exploration}. Springer, 2009,
  pp.~109--125.

\bibitem[ABK98]{amenta1998new}
\textsc{Amenta N., Bern M., Kamvysselis M.}:
\newblock A new voronoi-based surface reconstruction algorithm.
\newblock In \emph{Proceedings of the 25th annual conference on Computer
  graphics and interactive techniques} (1998), ACM, pp.~415--421.

\bibitem[ACDL00]{amenta2000simple}
\textsc{Amenta N., Choi S., Dey T.~K., Leekha N.}:
\newblock A simple algorithm for homeomorphic surface reconstruction.
\newblock In \emph{Proceedings of the sixteenth annual symposium on
  Computational geometry} (2000), ACM, pp.~213--222.

\bibitem[ACSd{\etalchar{*}}18]{url:optimaltransport}
\textsc{Alliez P., Cohen-Steiner D., deGoes F., Jamin C., Vigan I.}:
\newblock Optimal transportation reconstruction, 2018.

\bibitem[ACSTD07]{alliez2007voronoi}
\textsc{Alliez P., Cohen-Steiner D., Tong Y., Desbrun M.}:
\newblock Voronoi-based variational reconstruction of unoriented point sets.
\newblock In \emph{Symposium on Geometry processing} (2007), vol.~7,
  pp.~39--48.

\bibitem[AD15]{etudataset}
\textsc{Akimaliev M., Demirci M.~F.}:
\newblock Improving skeletal shape abstraction using multiple optimal
  solutions.
\newblock \emph{Pattern Recognition 48}, 11 (2015), 3504 -- 3515.

\bibitem[AM00]{althaus00polynomial}
\textsc{Althaus E., Mehlhorn K.}:
\newblock Tsp-based curve reconstruction in polynomial time.
\newblock In \emph{SODA '00: Proceedings of the eleventh annual ACM-SIAM
  symposium on Discrete algorithms} (Philadelphia, PA, USA, 2000), Soc. for
  Industr. and Appl. Math., pp.~686--695.

\bibitem[AMS00]{althaus00curve}
\textsc{Althaus E., Mehlhorn K., Schirra S.}:
\newblock Experiments on curve reconstruction.
\newblock In \emph{Proc. 2nd Workshop Algorithm Eng. Exper} (2000),
  pp.~103--114.

\bibitem[Aro98]{arora1998polynomial}
\textsc{Arora S.}:
\newblock Polynomial time approximation schemes for euclidean traveling
  salesman and other geometric problems.
\newblock \emph{Journal of the ACM (JACM) 45}, 5 (1998), 753--782.

\bibitem[ARZ05]{ahlvers2005model}
\textsc{Ahlvers U., Rajagopalan R., Z{\"o}lzer U.}:
\newblock Model-free face detection and head tracking with morphological hole
  mapping.
\newblock In \emph{2005 13th European Signal Processing Conference} (2005),
  IEEE, pp.~1--4.

\bibitem[Att97]{attali97regular}
\textsc{Attali D.}:
\newblock r-regular shape reconstruction from unorganized points.
\newblock In \emph{Symp. on Computational Geometry} (1997), pp.~248--253.

\bibitem[BB97]{bernardini1997sampling}
\textsc{Bernardini F., Bajaj C.~L.}:
\newblock Sampling and reconstructing manifolds using alpha-shapes.
\newblock \emph{Proc. 9th Canad. Conf. Comput. Geom.} (1997).

\bibitem[BLN{\etalchar{*}}13]{berger2013benchmark}
\textsc{Berger M., Levine J.~A., Nonato L.~G., Taubin G., Silva C.~T.}:
\newblock A benchmark for surface reconstruction.
\newblock \emph{ACM Transactions on Graphics (TOG) 32}, 2 (2013), 20.

\bibitem[Blu67]{blum67medialaxis}
\textsc{Blum H.}:
\newblock {A} {T}ransformation for {E}xtracting {N}ew {D}escriptors of {S}hape.
\newblock In \emph{Models for the Perception of Speech and Visual Form},
  Wathen-Dunn W., (Ed.). MIT Press, Cambridge, 1967, pp.~362--380.

\bibitem[BMR{\etalchar{*}}99]{bernardini1999ball}
\textsc{Bernardini F., Mittleman J., Rushmeier H., Silva C., Taubin G.}:
\newblock The ball-pivoting algorithm for surface reconstruction.
\newblock \emph{IEEE transactions on visualization and computer graphics 5}, 4
  (1999), 349--359.

\bibitem[Boi84a]{boissonat1984representing}
\textsc{Boissonat J.}:
\newblock Representing 2d and 3d shapes with the delaunay triangulation.
\newblock In \emph{InSeventh International Conference on Pattern Recognition
  (ICPR’84)} (1984).

\bibitem[Boi84b]{boissonnat1984geometric}
\textsc{Boissonnat J.-D.}:
\newblock Geometric structures for three-dimensional shape representation.
\newblock \emph{ACM Transactions on Graphics (TOG) 3}, 4 (1984), 266--286.

\bibitem[Boo79]{BOOKSTEIN197956}
\textsc{Bookstein F.~L.}:
\newblock Fitting conic sections to scattered data.
\newblock \emph{Computer Graphics and Image Processing 9}, 1 (1979), 56--71.

\bibitem[CBC{\etalchar{*}}01]{Carr:2001:RRO:383259.383266}
\textsc{Carr J.~C., Beatson R.~K., Cherrie J.~B., Mitchell T.~J., Fright W.~R.,
  McCallum B.~C., Evans T.~R.}:
\newblock Reconstruction and representation of 3d objects with radial basis
  functions.
\newblock In \emph{Proceedings of the 28th Annual Conference on Computer
  Graphics and Interactive Techniques} (New York, NY, USA, 2001), SIGGRAPH '01,
  ACM, pp.~67--76.

\bibitem[CFG{\etalchar{*}}05]{cheng2005curve}
\textsc{Cheng S.-W., Funke S., Golin M., Kumar P., Poon S.-H., Ramos E.}:
\newblock Curve reconstruction from noisy samples.
\newblock \emph{Computational Geometry 31}, 1-2 (2005), 63--100.

\bibitem[dALJ{\etalchar{*}}15]{10.1145/2732197}
\textsc{de~Ara\'{u}jo B.~R., Lopes D.~S., Jepp P., Jorge J.~A., Wyvill B.}:
\newblock A survey on implicit surface polygonization.
\newblock \emph{ACM Comput. Surv. 47}, 4 (May 2015).

\bibitem[Dey06]{dey2006curve}
\textsc{Dey T.~K.}:
\newblock \emph{Curve and surface reconstruction: algorithms with mathematical
  analysis}, vol.~23.
\newblock Cambridge University Press, 2006.

\bibitem[DGCSAD11]{degoes2011optimal}
\textsc{De~Goes F., Cohen-Steiner D., Alliez P., Desbrun M.}:
\newblock An optimal transport approach to robust reconstruction and
  simplification of 2d shapes.
\newblock In \emph{Computer Graphics Forum} (2011), vol.~30, Wiley Online
  Library, pp.~1593--1602.

\bibitem[DK99]{dey99curve}
\textsc{Dey T.~K., Kumar P.}:
\newblock A simple provable algorithm for curve reconstruction.
\newblock In \emph{Proc. 10th ACM-SIAM SODA '99} (1999), pp.~893--894.

\bibitem[DKWG08]{DUCKHAM20083224}
\textsc{Duckham M., Kulik L., Worboys M., Galton A.}:
\newblock Efficient generation of simple polygons for characterizing the shape
  of a set of points in the plane.
\newblock \emph{Pattern Recognition 41}, 10 (2008), 3224 -- 3236.

\bibitem[DMR99]{dey99conservative}
\textsc{Dey T.~K., Mehlhorn K., Ramos E.~A.}:
\newblock Curve reconstruction: Connecting dots with good reason.
\newblock \emph{In Proc. 15th ACM Symp. Comp. Geom 15} (1999), 229--244.

\bibitem[DT14]{duarte2014smoothness}
\textsc{Duarte P., Torres M.~J.}:
\newblock Smoothness of boundaries of regular sets.
\newblock \emph{Journal of mathematical imaging and vision} (2014), 1--8.

\bibitem[DW01]{dey01corners}
\textsc{Dey T.~K., Wenger R.}:
\newblock Reconstructing curves with sharp corners.
\newblock \emph{Computational Geometry 19}, 2-3 (2001), 89 -- 99.

\bibitem[DW02]{dey02gathang}
\textsc{Dey T.~K., Wenger R.}:
\newblock Fast reconstruction of curves with sharp corners.
\newblock \emph{Int. J. Comp. Geom. Appl. 12}, 5 (2002), 353 -- 400.

\bibitem[Ede98]{10.1007/BFb0054315}
\textsc{Edelsbrunner H.}:
\newblock Shape reconstruction with delaunay complex.
\newblock In \emph{LATIN'98: Theoretical Informatics} (Berlin, Heidelberg,
  1998), Lucchesi C.~L., Moura A.~V., (Eds.), Springer Berlin Heidelberg,
  pp.~119--132.

\bibitem[EKS83]{edelsbrunner83alpha}
\textsc{Edelsbrunner H., Kirkpatrick D.~G., Seidel R.}:
\newblock On the shape of a set of points in the plane.
\newblock \emph{IEEE Trans. Inf. Theor. IT-29}, 4 (1983), 551--559.

\bibitem[EM94]{edelsbrunner1994three}
\textsc{Edelsbrunner H., M{\"u}cke E.~P.}:
\newblock Three-dimensional alpha shapes.
\newblock \emph{ACM Transactions on Graphics (TOG) 13}, 1 (1994), 43--72.

\bibitem[FCOAS03]{Fleishman:2003:PPS:944020.944023}
\textsc{Fleishman S., Cohen-Or D., Alexa M., Silva C.~T.}:
\newblock Progressive point set surfaces.
\newblock \emph{ACM Trans. Graph. 22}, 4 (Oct. 2003), 997--1011.

\bibitem[FCOS05]{Fleishman:2005:RML:1073204.1073227}
\textsc{Fleishman S., Cohen-Or D., Silva C.~T.}:
\newblock Robust moving least-squares fitting with sharp features.
\newblock \emph{ACM Trans. Graph. 24}, 3 (July 2005), 544--552.

\bibitem[Fed59]{federer59curvature}
\textsc{Federer H.}:
\newblock Curvature measures.
\newblock \emph{Transactions of the American Mathematical Society 93}, 3
  (1959), pp. 418--491.

\bibitem[FMG94]{figueiredo94curve}
\textsc{Figueiredo L. H.~d., Mirandas~Gomes J.~d.}:
\newblock Computational morphology of curves.
\newblock \emph{Vis. Comp. 11}, 2 (1994), 105--112.

\bibitem[FMZB91]{FORSYTH1991130}
\textsc{Forsyth D., Mundy J.~L., Zisserman A., Brown C.~M.}:
\newblock Projectively invariant representations using implicit algebraic
  curves.
\newblock \emph{Image and Vision Computing 9}, 2 (1991), 130--136.

\bibitem[FR01]{funke01curve}
\textsc{Funke S., Ramos E.~A.}:
\newblock Reconstructing a collection of curves with corners and endpoints.
\newblock In \emph{Proceedings of the twelfth annual ACM-SIAM symposium on
  Discrete algorithms} (Philadelphia, PA, USA, 2001), SODA '01, Society for
  Industrial and Applied Math., pp.~344--353.

\bibitem[GDJ{\etalchar{*}}11]{5732742}
\textsc{Gheibi A., Davoodi M., Javad A., Panahi F., Aghdam M., Asgaripour M.,
  Mohades A.}:
\newblock Polygonal shape reconstruction in the plane.
\newblock \emph{Computer Vision, IET 5}, 2 (March 2011), 97 --106.

\bibitem[Gen90]{genoud90hamiltonian}
\textsc{Genoud T.}:
\newblock Etude du caract\`ere hamiltonien de delaunays al\'{e}atoires.
\newblock Travail de semestre, 1990.

\bibitem[GG07]{Guennebaud:2007:APS:1276377.1276406}
\textsc{Guennebaud G., Gross M.}:
\newblock Algebraic point set surfaces.
\newblock \emph{ACM Transactions on Graphics (TOG) 26}, 3 (2007), 23.

\bibitem[Gie99]{giesen99delaunay}
\textsc{Giesen J.}:
\newblock Curve reconstruction in arbitrary dimension and the traveling
  salesman problem.
\newblock In \emph{Proc. 8th DCGI '99} (1999), pp.~164--176.

\bibitem[Gol99]{gold99anticrust}
\textsc{Gold C.}:
\newblock Crust and anti-crust: a one-step boundary and skeleton extraction
  algorithm.
\newblock In \emph{Proc. of the 15th ann. Symp. on Computational geometry} (New
  York, NY, USA, 1999), SCG '99, ACM, pp.~189--196.

\bibitem[HDD{\etalchar{*}}92]{Hoppe:1992:SRU:142920.134011}
\textsc{Hoppe H., DeRose T., Duchamp T., McDonald J., Stuetzle W.}:
\newblock Surface reconstruction from unorganized points.
\newblock \emph{SIGGRAPH Comput. Graph. 26}, 2 (July 1992), 71--78.

\bibitem[Hiy09]{hiyoshi09optimize}
\textsc{Hiyoshi H.}:
\newblock Optimization-based approach for curve and surface reconstruction.
\newblock \emph{Comput. Aided Des. 41} (May 2009), 366--374.

\bibitem[HS09]{hoos2009empirical}
\textsc{Hoos H.~H., St{\"u}tzle T.}:
\newblock \emph{On the empirical scaling of run-time for finding optimal
  solutions to the traveling salesman problem}.
\newblock Tech. rep., Technical Report 17, University of British Columbia,
  Department of Computer Science, 2009.

\bibitem[IG16]{nonpoly1}
\textsc{{Iglesias} A., {Gálvez} A.}:
\newblock Cuckoo search with lévy flights for reconstruction of outline curves
  of computer fonts with rational bézier curves.
\newblock In \emph{2016 IEEE Congress on Evolutionary Computation (CEC)}
  (2016), pp.~2247--2254.

\bibitem[IGA18]{nonpoly2}
\textsc{Iglesias A., Galvez A., Avila A.}:
\newblock Immunological approach for full nurbs reconstruction of outline
  curves from noisy data points in medical imaging.
\newblock \emph{IEEE/ACM Trans. Comput. Biol. Bioinformatics 15}, 6 (Nov.
  2018), 1929–1942.

\bibitem[Jar77]{jarvis1977computing}
\textsc{Jarvis R.}:
\newblock Computing the shape hull of points in the plane.
\newblock In \emph{Proceedings of the IEEE Computing Society Conference on
  Pattern Recognition and Image Processing} (1977), New York, pp.~231--241.

\bibitem[JT92]{RNG}
\textsc{Jaromczyk J., Toussaint G.}:
\newblock Relative neighborhood graphs and their relatives.
\newblock \emph{Proceedings of the IEEE 80}, 9 (sep 1992), 1502 --1517.

\bibitem[Kaz05]{10.5555/1281920.1281931}
\textsc{Kazhdan M.}:
\newblock Reconstruction of solid models from oriented point sets.
\newblock In \emph{Proceedings of the Third Eurographics Symposium on Geometry
  Processing} (Goslar, DEU, 2005), SGP ’05, Eurographics Association,
  p.~73–es.

\bibitem[KBH06]{10.5555/1281957.1281965}
\textsc{Kazhdan M., Bolitho M., Hoppe H.}:
\newblock Poisson surface reconstruction.
\newblock In \emph{Proceedings of the Fourth Eurographics Symposium on Geometry
  Processing} (Goslar, DEU, 2006), SGP ’06, Eurographics Association,
  p.~61–70.

\bibitem[Kol08]{Kolluri:2008:PGM:1361192.1361195}
\textsc{Kolluri R.}:
\newblock Provably good moving least squares.
\newblock \emph{ACM Trans. Algorithms 4}, 2 (May 2008), 18:1--18:25.

\bibitem[KR85]{kirkpatrick85beta}
\textsc{Kirkpatrick D.~G., Radke J.~D.}:
\newblock A framework for computational morphology.
\newblock \emph{Computational Geometry} (1985), 217--248.

\bibitem[KR14]{khanna2014survey}
\textsc{Khanna K., Rajpal N.}:
\newblock Survey of curve and surface reconstruction algorithms from a set of
  unorganized points.
\newblock In \emph{Proceedings of the Third International Conference on Soft
  Computing for Problem Solving} (2014), Springer, pp.~451--458.

\bibitem[KTB07]{katz2007direct}
\textsc{Katz S., Tal A., Basri R.}:
\newblock Direct visibility of point sets.
\newblock In \emph{ACM SIGGRAPH 2007 papers}. 2007, pp.~24--es.

\bibitem[LC87]{10.1145/37402.37422}
\textsc{Lorensen W.~E., Cline H.~E.}:
\newblock Marching cubes: A high resolution 3d surface construction algorithm.
\newblock \emph{SIGGRAPH Comput. Graph. 21}, 4 (Aug. 1987), 163–169.

\bibitem[Lee00a]{lee2000curve}
\textsc{Lee I.-K.}:
\newblock Curve reconstruction from unorganized points.
\newblock \emph{Computer aided geometric design 17}, 2 (2000), 161--177.

\bibitem[Lee00b]{LEE2000161}
\textsc{Lee I.-K.}:
\newblock Curve reconstruction from unorganized points.
\newblock \emph{Computer Aided Geometric Design 17}, 2 (2000), 161 -- 177.

\bibitem[Len06]{lenz06curve}
\textsc{Lenz T.}:
\newblock How to sample and reconstruct curves with unusual features.
\newblock In \emph{Proceedings of the 22nd European Workshop on Computational
  Geometry (EWCG)} (Delphi, Greece, March 2006).

\bibitem[Lev98]{levin1998approximation}
\textsc{Levin D.}:
\newblock The approximation power of moving least-squares.
\newblock \emph{Mathematics of Computation of the American Mathematical Society
  67}, 224 (1998), 1517--1531.

\bibitem[Lev04]{Levin2004}
\textsc{Levin D.}:
\newblock \emph{Mesh-Independent Surface Interpolation}.
\newblock Springer Berlin Heidelberg, Berlin, Heidelberg, 2004, pp.~37--49.

\bibitem[Mat16]{mather2016foundations}
\textsc{Mather G.}:
\newblock \emph{Foundations of Sensation and Perception}.
\newblock Taylor \& Francis, 2016.

\bibitem[MKPM17]{METHIRUMANGALATH2017124}
\textsc{Methirumangalath S., Kannan S.~S., Parakkat A.~D., Muthuganapathy R.}:
\newblock Hole detection in a planar point set: An empty disk approach.
\newblock \emph{Computers \& Graphics 66} (2017), 124 -- 134.
\newblock Shape Modeling International 2017.

\bibitem[mpe02]{mpeg}
Mpeg-7 core experiment ce-shape-1 test set.
\newblock \url{https://cis.temple.edu/~latecki/TestData/mpeg7shapeB.tar.gz},
  2002.
\newblock [Online; accessed 19-October-2019].

\bibitem[MPM15]{METHIRUMANGALATH201590}
\textsc{Methirumangalath S., Parakkat A.~D., Muthuganapathy R.}:
\newblock A unified approach towards reconstruction of a planar point set.
\newblock \emph{Computers \& Graphics 51} (2015), 90 -- 97.
\newblock International Conference Shape Modeling International.

\bibitem[MPS08]{doi:10.1111/j.1467-8659.2008.01281.x}
\textsc{Manson J., Petrova G., Schaefer S.}:
\newblock Streaming surface reconstruction using wavelets.
\newblock \emph{Computer Graphics Forum 27}, 5 (2008), 1411--1420.

\bibitem[MTSM10]{mehra2010visibility}
\textsc{Mehra R., Tripathi P., Sheffer A., Mitra N.~J.}:
\newblock Visibility of noisy point cloud data.
\newblock \emph{Computers \& Graphics 34}, 3 (2010), 219--230.

\bibitem[NSW08]{niyogi2008finding}
\textsc{Niyogi P., Smale S., Weinberger S.}:
\newblock Finding the homology of submanifolds with high confidence from random
  samples.
\newblock \emph{Discrete \& Computational Geometry 39}, 1-3 (2008), 419--441.

\bibitem[NZ08]{nguyen08vicur}
\textsc{Nguyen T.~A., Zeng Y.}:
\newblock Vicur: A human-vision-based algorithm for curve reconstruction.
\newblock \emph{Robotics and Computer-Integrated Manufacturing 24}, 6 (2008),
  824 -- 834.
\newblock FAIM 2007, 17th International Conference on Flexible Automation and
  Intelligent Manufacturing.

\bibitem[OBA{\etalchar{*}}03]{Ohtake:2003:MPU:882262.882293}
\textsc{Ohtake Y., Belyaev A., Alexa M., Turk G., Seidel H.-P.}:
\newblock Multi-level partition of unity implicits.
\newblock \emph{ACM Trans. Graph. 22}, 3 (July 2003), 463--470.

\bibitem[OBS06]{Ohtake200615}
\textsc{Ohtake Y., Belyaev A., Seidel H.-P.}:
\newblock Sparse surface reconstruction with adaptive partition of unity and
  radial basis functions.
\newblock \emph{Graphical Models 68}, 1 (2006), 15 -- 24.
\newblock Special Issue on SMI 2004.

\bibitem[OBW87]{o1987connect}
\textsc{O'Rourke J., Booth H., Washington R.}:
\newblock Connect-the-dots: a new heuristic.
\newblock \emph{Computer Vision, Graphics, and Image Processing 39}, 2 (1987),
  258--266.

\bibitem[Ohr13]{url:connect2d}
\textsc{Ohrhallinger S.}:
\newblock An efficient algorithm for determining an aesthetic shape connecting
  unorganized 2d points.
\newblock \url{https://sourceforge.net/projects/connect2dlib/}, 2013.
\newblock [Online; accessed 12-September-2018].

\bibitem[Ohr16]{url:hnn-crust}
\textsc{Ohrhallinger S.}:
\newblock Half-nearest neighbor crust.
\newblock \url{https://github.com/stefango74/hnn-crust-sgp16}, 2016.
\newblock [Online; accessed 12-September-2018].

\bibitem[Ohr18a]{url:fitconnect}
\textsc{Ohrhallinger S.}:
\newblock Fitconnect.
\newblock \url{https://github.com/stefango74/fitconnect}, 2018.
\newblock [Online; accessed 12-September-2018].

\bibitem[Ohr18b]{url:stretchdenoise}
\textsc{Ohrhallinger S.}:
\newblock Stretchdenoise.
\newblock \url{https://github.com/stefango74/stretchdenoise}, 2018.
\newblock [Online; accessed 12-September-2018].

\bibitem[OM11]{ohrhallinger11operations}
\textsc{Ohrhallinger S., Mudur S.~P.}:
\newblock Interpolating an unorganized 2d point cloud with a single closed
  shape.
\newblock \emph{Computer-Aided Design 43}, 12 (2011), 1629--1638.

\bibitem[OM13]{ohrhallinger13connect2d}
\textsc{Ohrhallinger S., Mudur S.}:
\newblock An efficient algorithm for determining an aesthetic shape connecting
  unorganized 2d points.
\newblock In \emph{Comp. Graph. Forum} (2013), vol.~32, Wiley Online Library,
  pp.~72--88.

\bibitem[OMW16]{ohrhallinger2016hnn}
\textsc{Ohrhallinger S., Mitchell S.~A., Wimmer M.}:
\newblock Curve reconstruction with many fewer samples.
\newblock In \emph{Computer Graphics Forum} (2016), vol.~35, Wiley Online
  Library, pp.~167--176.

\bibitem[O'R81]{o1981polyhedra}
\textsc{O'Rourke J.}:
\newblock Polyhedra of minimal area as 3d object models.
\newblock In \emph{IJCAI} (1981), Citeseer, pp.~664--666.

\bibitem[OW18a]{ohrhallinger2018fitconnect}
\textsc{Ohrhallinger S., Wimmer M.}:
\newblock Fitconnect: Connecting noisy 2d samples by fitted neighbourhoods.
\newblock In \emph{Computer Graphics Forum} (2018), Wiley Online Library.

\bibitem[OW18b]{ohrhallinger2018stretchdenoise}
\textsc{Ohrhallinger S., Wimmer M.}:
\newblock Stretchdenoise: Parametric curve reconstruction with guarantees by
  separating connectivity from residual uncertainty of samples.
\newblock In \emph{Pacific Graphics} (2018), Wiley Online Library.

\bibitem[Par16]{url:crawl}
\textsc{Parakkat A.~D.}:
\newblock Crawl through neighbors.
\newblock
  \url{https://github.com/reproducibilitystamp/Crawl_through_Neighbors}, 2016.
\newblock [Online; accessed 12-September-2018].

\bibitem[Par18]{url:peeling}
\textsc{Parakkat A.~D.}:
\newblock Peeling the longest.
\newblock \url{https://github.com/amaldevp/Peeling_the_Longest}, 2018.
\newblock [Online; accessed 12-September-2018].

\bibitem[PM15a]{PEETHAMBARAN2015164}
\textsc{Peethambaran J., Muthuganapathy R.}:
\newblock A non-parametric approach to shape reconstruction from planar point
  sets through delaunay filtering.
\newblock \emph{Computer-Aided Design 62} (2015), 164 -- 175.

\bibitem[PM15b]{PEETHAMBARAN201562}
\textsc{Peethambaran J., Muthuganapathy R.}:
\newblock Reconstruction of water-tight surfaces through delaunay sculpting.
\newblock \emph{Computer-Aided Design 58} (2015), 62 -- 72.
\newblock Solid and Physical Modeling 2014.

\bibitem[PM16]{parakkat2016crawl}
\textsc{Parakkat A.~D., Muthuganapathy R.}:
\newblock Crawl through neighbors: A simple curve reconstruction algorithm.
\newblock In \emph{Computer Graphics Forum} (2016), vol.~35, Wiley Onl.
  Library, pp.~177--186.

\bibitem[PMM18]{parakkat2018peeling}
\textsc{Parakkat A.~D., Methirumangalath S., Muthuganapathy R.}:
\newblock Peeling the longest: A simple generalized curve reconstruction
  algorithm.
\newblock \emph{Computers \& Graphics 74} (2018), 191 -- 201.

\bibitem[PPT{\etalchar{*}}19]{doi:10.1111/cgf.13589}
\textsc{Peethambaran J., Parakkat A., Tagliasacchi A., Wang R., Muthuganapathy
  R.}:
\newblock Incremental labelling of voronoi vertices for shape reconstruction.
\newblock \emph{Computer Graphics Forum 0}, 0 (2019).

\bibitem[Pra87]{Vaughan87}
\textsc{Pratt V.}:
\newblock Direct least-squares fitting of algebraic surfaces.
\newblock \emph{SIGGRAPH Comput. Graph. 21}, 4 (Aug. 1987), 145–152.

\bibitem[Roh20]{webplot}
\textsc{Rohatgi A.}:
\newblock Webplotdigitizer: Version 4.3, 2020.

\bibitem[Rup93]{ruppert93lfs}
\textsc{Ruppert J.}:
\newblock A new and simple algorithm for quality 2-dimensional mesh generation.
\newblock In \emph{Proceedings of the fourth annual ACM-SIAM Symposium on
  Discrete algorithms} (Philadelphia, PA, USA, 1993), SODA '93, Soc. for
  Industr. and Appl. Math., pp.~83--92.

\bibitem[Rup14]{rupniewski2014curve}
\textsc{Rupniewski M.~W.}:
\newblock Curve reconstruction from noisy and unordered samples.
\newblock In \emph{3rd International Conference on Pattern Recognition
  Applications and Methods} (2014), SciTePress.

\bibitem[SJP10]{5521470}
\textsc{{Song} X., {Jüttler} B., {Poteaux} A.}:
\newblock Hierarchical spline approximation of the signed distance function.
\newblock In \emph{2010 Shape Modeling International Conference} (2010),
  pp.~241--245.

\bibitem[ST09]{stelldinger2009provably}
\textsc{Stelldinger P., Tcherniavski L.}:
\newblock Provably correct reconstruction of surfaces from sparse noisy
  samples.
\newblock \emph{Pattern Recognition 42}, 8 (2009), 1650--1659.

\bibitem[Ste08]{stelldinger2008topologically}
\textsc{Stelldinger P.}:
\newblock Topologically correct surface reconstruction using alpha shapes and
  relations to ball-pivoting.
\newblock In \emph{Pattern Recognition, 2008. ICPR 2008. 19th Int'l Conference
  on} (2008), IEEE, pp.~1--4.

\bibitem[{Tau}91]{Taubin}
\textsc{{Taubin} G.}:
\newblock Estimation of planar curves, surfaces, and nonplanar space curves
  defined by implicit equations with applications to edge and range image
  segmentation.
\newblock \emph{IEEE Transactions on Pattern Analysis and Machine Intelligence
  13}, 11 (1991), 1115--1138.

\bibitem[{Tau}93]{Taubin1}
\textsc{{Taubin} G.}:
\newblock An improved algorithm for algebraic curve and surface fitting.
\newblock In \emph{1993 (4th) International Conference on Computer Vision}
  (1993), pp.~658--665.

\bibitem[TO02]{Turk:2002:MIS:571647.571650}
\textsc{Turk G., O'Brien J.~F.}:
\newblock Modelling with implicit surfaces that interpolate.
\newblock \emph{ACM Trans. Graph. 21}, 4 (Oct. 2002), 855--873.

\bibitem[Tou88]{Toussaint}
\textsc{Toussaint G.}:
\newblock A graph-theoretical primal sketch.
\newblock \emph{Computational Morphology} (1988), 229--260.

\bibitem[TPM20]{THAYYIL2020101879}
\textsc{Thayyil S.~B., Parakkat A.~D., Muthuganapathy R.}:
\newblock An input-independent single pass algorithm for reconstruction from
  dot patterns and boundary samples.
\newblock \emph{Computer Aided Geometric Design 80} (2020), 101879.

\bibitem[TPM21]{THAYYIL2021101953}
\textsc{Thayyil S.~B., Peethambaran J., Muthuganapathy R.}:
\newblock A sampling type discernment approach towards reconstruction of a
  point set in r2.
\newblock \emph{Computer Aided Geometric Design 84} (2021), 101953.

\bibitem[Vel92]{veltkamp1992gamma}
\textsc{Veltkamp R.~C.}:
\newblock The $\gamma$-neighborhood graph.
\newblock \emph{Computational Geometry 1}, 4 (1992), 227--246.

\bibitem[Vel93]{veltkamp19933d}
\textsc{Veltkamp R.~C.}:
\newblock 3d computational morphology.
\newblock In \emph{Computer Graphics Forum} (1993), vol.~12, Wiley Online
  Library, pp.~115--127.

\bibitem[WYZ{\etalchar{*}}14]{wang2014robust}
\textsc{Wang J., Yu Z., Zhang W., Wei M., Tan C., Dai N., Zhang X.}:
\newblock Robust reconstruction of 2d curves from scattered noisy point data.
\newblock \emph{Computer-Aided Design 50} (2014), 27--40.

\bibitem[ZNYL08]{zeng08distance}
\textsc{Zeng Y., Nguyen T.~A., Yan B., Li S.}:
\newblock A distance-based parameter free algorithm for curve reconstruction.
\newblock \emph{Comput. Aided Des. 40}, 2 (2008), 210--222.

\end{thebibliography}
\end{document}